\ifx\fiverm\undefined
	\newfont\fiverm{cmr5}
\fi
\input prepictex
\input pictex
\input postpictex
\catcode`\@=11
\@input{picmore.tex}

\documentstyle[11pt]{article}

\def\labelmark{}

\def\void{}
\newenvironment{formula}[1]{\def\labelname{#1}
\ifx\void\labelname\def\junk{\begin{displaymath}}
\else\def\junk{\begin{equation}\label{\labelname}}\fi\junk}%
{\ifx\void\labelname\def\junk{\end{displaymath}}
\else\def\junk{\end{equation}}\fi\junk\labelmark\def\labelname{}}

{\ifx\void\labelname\def\junk{\end{array}\end{displaymath}}
\else\def\junk{\end{array}\right.\end{equation}}
\fi\junk\labelmark\def\labelname{}\def\junk{}
}

\newcommand{\beq}{\begin{formula}}
\newcommand{\eeq}{\end{formula}}

\newcommand{\beqa}{\begin{eqnarray}}
\newcommand{\eeqa}{\end{eqnarray}}
\newcommand{\eq}[1]{(\ref{#1})}
\newcommand{\nn}{\nonumber}
\newcommand{\tlim}[1]{\stackrel{{#1} \rightarrow 0}{\longrightarrow}}

\newcommand{\rcov}[2]{{\stackrel{\leftarrow}{D}}_{#1}
\hskip-5pt{}^{#2}}
\newcommand{\define}{\stackrel{\mbox{def.}}{=}}
\newcommand{\ra}{\rightarrow}
\newcommand{\der}{\partial}
\newcommand{\fder}{\delta}
\newcommand{\gmnrs}{\delta^{\mu'\nu'}_{\mu\phantom{{}'}\nu}
\delta^{\rho'\sigma'}_{\rho\phantom{{}'}\sigma}}
\newcommand{\dand}{\hskip-6pt &}
\newcommand{\NP}[1]{ {\it Nucl.~Phys.} {\bf #1}}
\newcommand{\NPS}[1]{ {\it Nucl.~Phys.~(Proc.~Suppl.)} {\bf #1}} 
\newcommand{\PL}[1]{ {\it Phys.~Lett.} {\bf #1}}

\newcommand{\PR}[1]{ {\it Phys.~Rev.} {\bf #1}}
\newcommand{\PRL}[1]{ {\it Phys.~Rev.~Lett.} {\bf #1}}

\newcommand{\AP}[1]{ {\it Ann.~Phys.} {\bf #1}}

\newcommand{\ACT}[1]{ {\it Acta~Phys.~Polon.} {\bf #1}}
\newcommand{\EPJ}[1]{ {\it Europ.~Phys.~J.} {\bf #1}}
\newcommand{\TMP}[1]{ {\it Theor.~Math.~Phys.} {\bf #1}}

\begin{document}
\begin{titlepage}
\setcounter{page}{0}
\renewcommand{\thefootnote}{\fnsymbol{footnote}}

\begin{flushright}
HD-THEP-98-64\\
hep-th/9812229\\
\end{flushright}

\vspace{5 mm}
\begin{center}
{\Large\bf World-line approach to the Bern-Kosower formalism \\
in two-loop Yang-Mills theory} 

\vspace{10 mm}
{\bf Haru-Tada Sato$^{}$
\footnote{E-mail: sato@thphys.uni-heidelberg.de }}
\,\sc{and}\,
{\bf Michael G. Schmidt$^{}$
\footnote{E-mail: m.g.schmidt@thphys.uni-heidelberg.de}}

\vspace{5mm}

\vspace{10mm}

{\it $^{}$ Institut f{\"u}r Theoretische Physik\\
Universit{\"a}t Heidelberg\\
Philosophenweg 16, D-69120 Heidelberg, Germany}
\end{center}

\vspace{10mm}
\begin{abstract}
Based on the world-line formalism with a sewing method, 
we derive the Yang-Mills effective action in a form useful to 
generate the Bern-Kosower-type master formulae for gluon scattering 
amplitudes at the two-loop level. It is shown that four-gluon 
($\Phi^4$ type sewing) contributions can be encapsulated in the 
action with three-gluon ($\Phi^3$ type) vertices only, the 
total action thus becoming a simple expression. We then derive 
a general formula for a two-loop Euler-Heisenberg type action in 
a pseudo-abelian $su(2)$ background. 
The ghost loop and fermion loop cases are also studied. 
\end{abstract}

\vfill

\begin{flushleft}
\end{flushleft}
\end{titlepage}
\setcounter{footnote}{0}
\renewcommand{\thefootnote}{\arabic{footnote}}
\renewcommand{\theequation}{\thesection.\arabic{equation}}
\section{Introduction}\label{sec1}
\setcounter{section}{1}
\setcounter{equation}{0}
\indent

More than a decade ago, the analogy between first quantized approach 
in string theory and world-line representation in field theory was 
pointed out in the $\phi^3$ theory effective action~\cite{zet1}, and 
a similar approach was considered for Yang-Mills theory~\cite{zet2}. 
Since then, the relation of string theory to quantum field 
theory has been studied intensively for the particular purpose of 
obtaining field theory scattering amplitudes in a remarkably simple 
way~\cite{BK}-\cite{sYM}. String theory organizes scattering 
amplitudes in a compact form (by virtue of the conformal symmetry on 
the world-sheet), and field theory, as a singular limit of string 
theory, inherits this useful feature, by which the summation 
of Feynman diagrams is already installed without need of performing 
loop integrals and the Dirac traces. In particular, Bern and 
Kosower derived a set of simple rules for one-loop gluon scattering 
amplitudes through analyzing the field theory limit of a heterotic 
string theory~\cite{BK}. The rules turned out to correspond to a 
subtle combination of the background and Gervais-Neveu 
gauges~\cite{FTL}. They are applied to five gluon 
amplitudes~\cite{five,upto5}, quantum gravity~\cite{gra} and super 
Yang-Mills theories~\cite{sYM}. The Bern-Kosower (BK) rules are also 
derivable from bosonic string theory~\cite{bose}, and the concrete 
identification of a corner of moduli space with each Feynman diagram 
and its divergence is verified~\cite{paolo}. 

The completion of Feynman diagram summation means gauge invariance. 
It is well-known that the Feynman diagram calculation splits a gauge 
invariant amplitude into non-invariant terms, and this causes a 
cancellation between divergent diagrams (gauge cancellation), which 
brings a serious problem especially with numerical computation. In 
the Bern-Kosower formalism we do not have this problem, since 
the only divergence appears from the final integration of a 
universal master formula. The master formula does not depend on the 
simplicity of specific scatterings with small number of external 
legs. (We want to keep this universality as much as possible when 
considering multi-loop generalization.) 
Hence the BK formalism has a great deal of potential to renovate 
the computational technique and efficiency in quantum field theory.

The BK rules for one-loop cases are also attainable directly 
(without making use of string theory) in terms of the world-line 
method in quantum field theory~\cite{St,poly}. In this case, we 
have to evaluate an effective action in some particular form, which 
is a path integral for a one-dimensional quantum mechanical action 
(world-line action), using the proper time and background field 
methods as well. Then expanding the background field as a sum of 
Fourier plane wave modes, we get the same kind of objects that are 
called the vertex operators in string theory. One particle 
irreducible (1PI) Green functions can be obtained as multi-integrals 
of the master formula, which is a correlation function evaluated 
by Wick's contraction with the two-point correlator (world-line 
Green function) determined from the world-line action. It is very 
interesting that this kind of vertex operator technique resembles 
string theory calculations, and that all Feynman diagrams are 
consequently contained in a single master formula like string 
theory amplitudes. In fact, various field theory examples can be 
understood from this viewpoint: Photon splitting~\cite{const}, 
axion decay in a constant magnetic field~\cite{ax}, and Yukawa 
interactions~\cite{yukawa} up to some finite values of $N$ 
(the number of external legs) are explicitly verified; for 
photon scattering and $\phi^3$ theory, the equivalence is 
formally proven up to the two-loop order with an arbitrary value 
of $N$~\cite{combi}. This formalism is also useful for a manifestly 
covariant calculation of the effective action~\cite{EA}, and for 
decompositions into gauge invariant partial amplitudes~\cite{part}.  

{}~Furthermore, since the first proposal of a multi-loop 
generalization of the Bern-Kosower formalism~\cite{SSphi}, various 
steps in this direction have been made~\cite{combi}-\cite{KS} 
--- mostly investigated in $\phi^3$ theory~\cite{combi}-\cite{RS1} 
and in spinor/scalar QED~\cite{SSqed}-\cite{RSS}. A few 
preliminary studies for QCD have also been 
performed~\cite{HTS2,vacuum,Kaj}. Generally speaking, we need some 
new types of master formulae, depending on the places where 
external legs are inserted. The multi-loop combinatorial problem, 
which is how to combine the master formulae of different types, 
is solved in the cases of neutral $\phi^3$ theory and scalar/spinor 
QED at the two-loop order ~\cite{combi}. The other new participants 
are the multi-loop world-line Green 
functions~\cite{SSphi}-\cite{RS1,RSS} and path integral 
normalizations. In $\phi^3$ theory, they are determined 
from the string theory side as well~\cite{RS1}-\cite{RS2}. 

Now, all ingredients seem ready to be generalized to the gluon 
scattering case at the two-loop level within the world-line 
framework, since basic features are almost in common with the 
above cases, up to the four-gluon interaction. On first thoughts, 
we might have only to insert a path integral representation of a 
gluon propagator~\cite{RSS} in a one-loop gluon diagram. However, 
things are not really straightforward. The insertion of a 
propagator destroys the simple trace structure, and we hence have 
to find out an alternative expression of the trace-log formula. In 
addition, in order to determine precise multi-loop combinatorics, 
we have to construct the two-loop analogue of the trace-log formula 
in a systematic way.{}~Fortunately we already have a suitable 
technique for these purposes. It all can be done by introducing an 
auxiliary field representing a quadratic term of the internal 
quantum field~\cite{combi}. Roughly speaking, the auxiliary field 
plays the role of an adhesive to glue the inserted propagator. We 
shall show more details how to exactly realize this idea later on. 
Once having a precise formula for the two-loop effective action, 
we expect that the substitution by a sum of all plane waves 
\beq{FT}
A_\mu^a {\hat\lambda}^a \quad\ra\quad 
g\sum_{i=1}^N {\hat\lambda}^{a_i}\epsilon^i_\mu \exp[ik_i\cdot x]
\eeq 
will yield correct combinatorics even in the multi-loop cases. 
This is actually verified in the $\phi^3$ theory case up to the 
two-loop order~\cite{combi}. Regarding this point, we shall confine 
ourselves to discuss a general prescription at the present stage. 

In this paper, we present a derivation of the world-line formulae 
for the two-loop effective action mainly in pure Yang-Mills theory. 
In Section~\ref{sec2}, starting with the background field Lagrangian 
together with the auxiliary field method mentioned above, we set up 
the sewing between a loop and a propagator so as to generate the 
two-loop analogue of a trace-log formula, which consists of four 
types of sewing. We also address more general sewing rules for a 
multi-loop construction. In Section~\ref{sec3}, we explicitly 
perform the sewing procedures at the level of world-line path 
integral representations. Briefly observing a conjecture in 
Section~\ref{sec4}, we then unify three of the four types into a 
single expression in Section~\ref{sec5}. In Section~\ref{sec6}, 
we verify this fact in the $su(2)$ pseudo-abelian case, and derive 
a general formula for the effective action in a constant 
(pseudo-abelian) background field (Euler-Heisenberg type action). 
We also include short remarks on the ghost loop case in 
Section~\ref{sec7} and on the fermion loop case in 
Section~\ref{sec8}. Conclusions and discussions are in 
Section~\ref{sec9}. In Appendix A, we attach two kinds of 
non-world-line calculations for comparison. One is solely based 
on the auxiliary field method (without the use of world-line 
representations), and the other is based on the usual field 
theory method. It is shown that the results of the main text 
perfectly coincide with those obtained by these two different 
methods. In Appendix B we show an outline of the method how to 
obtain (pure Yang-Mills) $N$-point amplitudes, and in Appendix C 
computational details of two-loop gluon 
world-line Green function are presented. 

\section{The two-loop analogue of the trace-log formula}\label{sec2}
\setcounter{section}{2}
\setcounter{equation}{0}
\indent


In this section, we derive the two-loop analogue of the trace-log 
formula for the pure Yang-Mills effective action, starting with the 
following background field Lagrangian~\cite{BGF} (for the moment we 
include the fermion and ghost Lagrangians as well):
\beqa
{\cal L}=&-&{1\over4}(F^a_{\mu\nu})^2 
-{1\over2}Q^a_\mu(\Delta^{-1})^{ab}_{\mu\nu}Q^b_\nu 
-F^a_{\mu\nu}D^{ac}_\mu Q^c_\nu 
-f^{abc}(D_\mu Q_\nu)^a Q^b_\mu Q^c_\nu \nn\\
&-&{1\over4}(f^{abc}Q^b_\mu Q^c_\nu)^2 + 
{\bar c}^a (D^2)^{ab}c^b + 
f^{abc}(D_\mu{\bar c})^ac^b Q^c_\mu \nn\\
&+&{\bar\psi}_ii\gamma^\mu({\hat D}_\mu-iQ^a_\mu{\hat\lambda}^a)_{ij}
\psi_j\ ,\label{lag}
\eeqa
where $Q^a_\mu$ is the quantum gauge field, and $D_\mu$ is 
the covariant derivative w.r.t. a background gauge field $A^a_\mu$, 
whose field strength is denoted by $F^a_{\mu\nu}$. The (non-) hat 
notations stand for the (adjoint) fundamental representations, and 
$[{\hat\lambda}^a,{\hat\lambda}^b]=if^{abc}{\hat\lambda}^c$. 
The propagator $\Delta$ in a general background gauge is given by 
\beq{*}
\Delta^{ab}_{\mu\nu} =[ 
-g_{\mu\nu}D^2+(1-{1\over\xi})D_\mu D_\nu
+2F^c_{\mu\nu}f^{abc}]^{-1} \ ,
\eeq
however we choose the Feynman gauge $\xi=1$ in this paper. 
Here are some notational remarks. We often put the Lorentz 
indices upside down just by reason of typography, and abbreviate 
the indices and space-time integrals which can be understood 
from the same/corresponding terms written beforehand. We also 
follow the convention $(X_{\mu\nu})^2\equiv X_{\mu\nu}X^{\mu\nu}$ 
for an arbitrary tensor, and the Lorentz and color summations 
are implicit in doubly appearing indices as usual. 

Basically, we follow the same method as we developed in $\phi^3$ 
theory (The details can be found in Section 3 of Ref.~\cite{combi}). 
We introduce the three sets of auxiliary fields regarding the 
quantum gauge field, ghost and quark parts. Inserting the 
following identities in the Lagrangian \eq{lag}:
\beqa
\delta(B-fQQ)&\equiv&\int{\cal D}\alpha^a_{\mu\nu}\,
\exp\Bigl[i\int d^Dx(B^a_{\mu\nu}-f^{abc}Q^b_\mu Q^c_\nu)
\alpha^a_{\mu\nu}\,\Bigr] \ ,\\
\delta(C-f(Dc)c)&\equiv&\int{\cal D}\beta^a_\mu\,
\exp\Bigl[i\int d^Dx(C^a_\mu-f^{abc}(D_\mu{\bar c})^bc^c)
\beta^a_\mu\,\Bigr] \ , \\
\delta(E-{\bar\psi}\gamma^\mu{\hat\lambda}^a\psi) &\equiv&
\int{\cal D}\epsilon^a_\mu\,
\exp\Bigl[i\int d^Dx(E^a_\mu - {\bar\psi}_i\gamma_\mu
({\hat\lambda}^a)_{ij}\psi_j)\epsilon^a_\mu
\,\Bigr]\ ,
\eeqa
we obtain the generating functional in the form 
\beqa
Z[A]&=&Z_0\int{\cal D}B^a_{\mu\nu}{\cal D}\alpha^a_{\mu\nu}
{\cal D}C^a_\mu{\cal D}\beta^a_\mu
{\cal D}E^a_\mu{\cal D}\epsilon^a_\mu \,[\mbox{Det}\,
{\bar\Delta}^{-1}]^{-1/2} \mbox{Det}[(D^2)^{ac}
-D^{ab}_\mu \beta^d_\mu f^{bdc}]\nn\\
&\times& \mbox{Det}
[i\gamma^\mu({\hat D}_\mu+i\epsilon^a_\mu \lambda^a)_{ij}]\,
\exp[{i\over2}\int d^Dy_1 d^D y_2 
J^a_\mu(y_1){\bar\Delta}^{ab}_{\mu\nu}(y_1,y_2)J^b_\nu(y_2)] \nn\\
&\times&\exp[i\int d^Dx (-{1\over4}(B^a_{\mu\nu})^2
+\alpha^a_{\mu\nu}B^a_{\mu\nu} +\beta^a_\mu C^a_\mu 
+\epsilon^a_\mu E^a_\mu )]\ ,\label{action} 
\eeqa
where
\beqa
&&J^a_\nu[B,C,E,F]=D^{ac}_\mu F^c_{\mu\nu} 
+ D^{ac}_\mu B^c_{\mu\nu} + C^a_\nu +E^a_\nu\ ,\label{source}\\
&&({\bar\Delta}^{-1})^{ab}_{\mu\nu} = (\Delta^{-1})^{ab}_{\mu\nu}
+2f^{abc}\alpha^c_{\mu\nu} \ ,\\
&&Z_0=\exp[\,i\int -{1\over4}(F^a_{\mu\nu})^2\, ]\ .
\eeqa
To perform the $B$, $C$ and $E$ integrals, we apply the following 
(general) formula for a function of $B$ and $\alpha$ 
\beq{*}
\int{\cal D}\alpha{\cal D}Bf(iB,\alpha)e^{i\alpha B}=
\int{\cal D}\alpha{\cal D}Bf({\fder\over\fder\alpha},\alpha)e^{i\alpha B}
=\int{\cal D}\alpha f({\fder\over\fder\alpha},\alpha)\delta(\alpha)\ ,
\eeq 
where the $\alpha$ differentiation acts on the $\delta$-function 
$\delta(\alpha)$. Keeping the partial integrations for the 
$\delta$-function in mind, one can see that the integrations of 
these $\delta$-functions then lead to the following replacements 
in the functional~\eq{action}:
\beq{*}
B^a_{\mu\nu}\ra {i\over2}
{\fder\over\fder\alpha^a_{\mu\nu}}\Bigr|_{\alpha=0}\ ,\qquad
C^a_\mu \ra i{\fder\over\fder\beta^a_\mu}\Bigr|_{\beta=0}\ ,\qquad
E^a_\mu\ra i{\fder\over\fder\epsilon^a_\mu}\Bigr|_{\epsilon=0}\ ,
\eeq
where ${1\over2}$ derives from the anti-symmetric nature of 
the $B$ field. Then the functional differentiations become to 
act on all $\alpha$, $\beta$ and $\epsilon$ fields. 
Removing irrelevant parts of the effective action, we obtain 
\beqa
Z[A] &\approx& \exp[-{1\over2}\mbox{Trln}{\bar\Delta}^{-1}]
\exp[\mbox{Trln}(D^2-D\beta f)] 
\exp[\mbox{Trln}(\gamma^\mu(i{\hat D}_\mu-
\epsilon^a_\mu{\hat\lambda}^a)] \nn\\
&\times&\exp\Bigl[\,{i\over2}(-{i\over2}{\fder\over\fder\alpha}D
+i{\fder\over\fder\beta}+i{\fder\over\fder\epsilon} ){\bar\Delta}
(-{i\over2}{\fder\over\fder\alpha}D+i{\fder\over\fder\beta} 
+i{\fder\over\fder\epsilon} )\,\Bigr]\nn\\
&\times&\exp[\,{i\over16}
{\fder\over\fder\alpha}1{\fder\over\fder\alpha}\,]
\Bigr|_{\alpha=\beta=\epsilon=0} \ .\label{ZA}
\eeqa
Since all the functional differentiations act on every $\alpha$ 
field etc., the ordering of the exponential objects is not important. 
The covariant derivatives appearing with the $\alpha$ differentiation 
should be understood as partially integrated ones (thus acting on the 
propagator) in order to get rid of acting on the functional derivatives. 
Thus the pure Yang-Mills part of $Z[A]$ reads 
\beq{pureac}
iZ^{gluon}= \exp[({i\over2})^3
D\fder_\alpha {\bar\Delta} D\fder_\alpha]\exp[{i\over16}\fder_\alpha
\fder_\alpha]
\exp[-{1\over2}\mbox{Trln}{\bar\Delta}^{-1}] \Bigr|_{\alpha=0}\ ,
\eeq
where $\delta_\alpha$ is the abbreviation of $\fder/\fder\alpha$. 
The third exponential object in Eq.\eq{pureac} generates loops 
including the one-loop effective action, and the first one the 
propagator insertions to produce multi-loop diagrams by three-gluon 
interactions. The second one corresponds to four-gluon interactions. 
One can see the similarity of this gluon action \eq{pureac} to the 
following $\phi^3$ theory action (massless, Euclidean)~\cite{combi}:
\beq{phi3}
Z[{\bar\phi}] = 
\exp[\,-{g^2\over2(3!)^2}\fder_\alpha
(-\der^2+g{\bar\phi}+2i\alpha)^{-1}\fder_\alpha \,]
\exp\Bigl[ -{1\over2}
\mbox{Tr}\ln(-\der^2+g{\bar\phi}+2i\alpha)\Bigr]
\Bigr|_{\alpha=0}\ . 
\eeq

Now we extract the two-loop (1PI) parts from the pure Yang-Mills 
generating functional \eq{pureac} as shown in Fig. 1. The 
two-loop effective action comprises the following four types:
\beqa
i\Gamma_1&=&-{1\over2}\int d^Dy_1d^Dy_2
({i\over2})^3 \fder_\alpha D {\bar\Delta} \fder_\alpha D
\mbox{Trln}{\bar\Delta}^{-1}
\Bigr|_{\alpha=0}\ ,\label{iW1}\\
i\Gamma_2&=&\int d^Dy_1d^Dy_2({i\over2})^3 \fder_\alpha D 
{\bar\Delta} \fder_\alpha D\Bigr|^{1PI}_{\alpha=0}\ ,\label{iW2} \\
i\Gamma_3^{(1)}&=&-{1\over2}\int d^Dy_1
{i\over16}\fder_\alpha\fder_\alpha
\mbox{Trln}{\bar\Delta}^{-1}
\Bigr|_{\alpha=0}\ ,\label{iW31}\\
i\Gamma_3^{(2)}&=&{1\over2}\int d^Dy_1 
{i\over16}\fder_\alpha\fder_\alpha
(-{1\over2}\mbox{Trln}{\bar\Delta}^{-1})^2 
\Bigr|^{connected}_{\alpha=0}\ ,
\label{iW32}
\eeqa
where we have revived the omitted space-time integrations, 
and one may of course insert $1=\int d^Dy_2\delta(y_1-y_2)$ into 
the $\delta_\alpha$ square terms (acting on the same point) 
in Eqs.\eq{iW31} and \eq{iW32}. In $\Gamma_2$ and $\Gamma^{(2)}_3$, 
we have to extract the 1PI pieces from the naive $\alpha$ 
differentiations. These will be explained in Section 3. 
We shall refer to the diagram (a) in Figure 1 as the propagator 
insertion type, (b) the double folding type, and to the rests 
as the shrinkage types or the eight figure diagrams. 

%
%
\vspace{8mm}
\begin{minipage}[t]{14.5cm} 
%
%
\font\thinlinefont=cmr5
\begingroup\makeatletter\ifx\SetFigFont\undefined
\def\x#1#2#3#4#5#6#7\relax{\def\x{#1#2#3#4#5#6}}%
\expandafter\x\fmtname xxxxxx\relax \def\y{splain}%
\ifx\x\y   
\gdef\SetFigFont#1#2#3{%
  \ifnum #1<17\tiny\else \ifnum #1<20\small\else
  \ifnum #1<24\normalsize\else \ifnum #1<29\large\else
  \ifnum #1<34\Large\else \ifnum #1<41\LARGE\else
     \huge\fi\fi\fi\fi\fi\fi
  \csname #3\endcsname}%
\else
\gdef\SetFigFont#1#2#3{\begingroup
  \count@#1\relax \ifnum 25<\count@\count@25\fi
  \def\x{\endgroup\@setsize\SetFigFont{#2pt}}%
  \expandafter\x
    \csname \romannumeral\the\count@ pt\expandafter\endcsname
    \csname @\romannumeral\the\count@ pt\endcsname
  \csname #3\endcsname}%
\fi
\fi\endgroup
\mbox{\beginpicture
\setcoordinatesystem units <0.45000cm,0.45000cm>
\unitlength=0.45000cm
\linethickness=1pt
\setplotsymbol ({\makebox(0,0)[l]{\tencirc\symbol{'160}}})
\setshadesymbol ({\thinlinefont .})
\setlinear
%
%
\linethickness=0.500pt
\setplotsymbol ({\thinlinefont .})
\circulararc 107.171 degrees from 14.531 21.770 center at 11.910 19.485
\circulararc 94.304 degrees from 11.917 20.743 center at 14.496 23.588
%
%
\linethickness=0.500pt
\setplotsymbol ({\makebox(0,0)[l]{\tencirc\symbol{'175}}})
\put{\makebox(0,0)[l]{\circle*{ 0.318}}} at 11.970 20.701
\put{\makebox(0,0)[l]{\circle*{ 0.318}}} at 14.541 21.780
%
%
\linethickness=0.500pt
\setplotsymbol ({\thinlinefont .})
\ellipticalarc axes ratio  1.943:1.943  360 degrees 
	from 32.042 20.955 center at 30.099 20.955
\ellipticalarc axes ratio  1.909:1.909  360 degrees 
	from 36.580 20.987 center at 34.671 20.987
%
%
\linethickness=0.500pt
\setplotsymbol ({\thinlinefont .})
\ellipticalarc axes ratio  2.529:2.529  360 degrees 
	from  6.657 20.923 center at  4.128 20.923
%
%
\linethickness=0.500pt
\setplotsymbol ({\makebox(0,0)[l]{\tencirc\symbol{'175}}})
\put{\makebox(0,0)[l]{\circle*{ 0.318}}} at  4.096 23.019
\put{\makebox(0,0)[l]{\circle*{ 0.318}}} at  4.096 18.891
%
%
\linethickness=1pt
\setplotsymbol ({\thinlinefont .})
\ellipticalarc axes ratio  0.135:0.135  360 degrees 
	from 23.012 20.936 center at 22.877 20.936
%
%
\linethickness=1pt
\setplotsymbol ({\thinlinefont .})
\ellipticalarc axes ratio  0.135:0.135  360 degrees 
	from 32.520 20.987 center at 32.385 20.987
%
%
\linethickness=0.500pt
\setplotsymbol ({\makebox(0,0)[l]{\tencirc\symbol{'175}}})
\putrule from  4.096 23.008 to  4.096 18.828
%
%
\linethickness=1pt
\setplotsymbol ({\thinlinefont .})
\plot  3.905 23.717  4.381 23.241 /
\plot  4.350 23.717  3.873 23.241 /
\plot  4.381 18.605  3.905 18.129 /
\plot  3.873 18.605  4.350 18.129 /
%
%
\linethickness=0.500pt
\setplotsymbol ({\makebox(0,0)[l]{\tencirc\symbol{'175}}})
\putrule from  8.954 21.273 to 17.526 21.273
%
%
\linethickness=1pt
\setplotsymbol ({\thinlinefont .})
\plot 15.208 21.050 14.760 21.499 /
\plot 14.764 21.050 15.212 21.499 /
\plot 11.557 21.082 12.006 21.531 /
\plot 11.970 21.082 11.521 21.531 /
%
%
\linethickness=1pt
\setplotsymbol ({\thinlinefont .})
\plot 22.860 21.273 22.860 21.946 /
\plot 22.543 21.558 23.216 21.558 /
\plot 22.892 19.971 22.892 20.644 /
\plot 22.574 20.320 23.247 20.320 /
%
%
%
\linethickness=1pt
\setplotsymbol ({\thinlinefont .})
\plot 31.814 20.765 32.262 21.213 /
\plot 32.262 20.760 31.814 21.209 /
\plot 32.988 20.733 32.540 21.181 /
\plot 32.575 20.733 33.024 21.181 /
%
%
\linethickness=0.500pt
\setplotsymbol ({\thinlinefont .})
\plot 22.828 21.558 	22.751 21.653
	22.675 21.744
	22.602 21.831
	22.530 21.914
	22.460 21.994
	22.392 22.069
	22.325 22.140
	22.260 22.208
	22.133 22.332
	22.010 22.442
	21.892 22.538
	21.777 22.621
	21.665 22.691
	21.555 22.749
	21.445 22.794
	21.336 22.828
	21.227 22.852
	21.117 22.864
	21.006 22.867
	20.892 22.860
	20.803 22.849
	20.715 22.834
	20.627 22.815
	20.540 22.793
	20.453 22.766
	20.368 22.737
	20.283 22.703
	20.200 22.667
	20.118 22.626
	20.038 22.583
	19.959 22.537
	19.883 22.487
	19.808 22.434
	19.736 22.378
	19.666 22.320
	19.599 22.258
	19.535 22.194
	19.473 22.127
	19.415 22.058
	19.360 21.986
	19.309 21.912
	19.261 21.835
	19.217 21.756
	19.176 21.675
	19.140 21.591
	19.109 21.506
	19.081 21.419
	19.059 21.329
	19.041 21.238
	19.028 21.146
	19.021 21.051
	19.018 20.955
	19.021 20.858
	19.030 20.763
	19.044 20.670
	19.062 20.578
	19.086 20.488
	19.114 20.400
	19.147 20.315
	19.184 20.231
	19.226 20.150
	19.271 20.070
	19.320 19.994
	19.373 19.919
	19.429 19.847
	19.489 19.778
	19.551 19.711
	19.617 19.647
	19.685 19.586
	19.756 19.527
	19.829 19.472
	19.905 19.419
	19.983 19.370
	20.062 19.323
	20.143 19.280
	20.226 19.241
	20.310 19.204
	20.396 19.171
	20.482 19.142
	20.569 19.116
	20.657 19.094
	20.746 19.075
	20.834 19.061
	20.923 19.050
	21.037 19.043
	21.148 19.046
	21.258 19.059
	21.367 19.083
	21.476 19.117
	21.585 19.163
	21.695 19.221
	21.807 19.291
	21.922 19.373
	22.041 19.469
	22.163 19.579
	22.226 19.639
	22.290 19.703
	22.356 19.771
	22.423 19.842
	22.491 19.917
	22.561 19.996
	22.633 20.079
	22.707 20.166
	22.782 20.257
	22.860 20.352
	/
\plot 22.924 20.384 	23.001 20.288
	23.076 20.197
	23.150 20.110
	23.222 20.027
	23.292 19.948
	23.360 19.873
	23.427 19.802
	23.492 19.734
	23.619 19.610
	23.741 19.500
	23.860 19.403
	23.974 19.321
	24.087 19.251
	24.197 19.193
	24.307 19.148
	24.415 19.113
	24.524 19.090
	24.634 19.078
	24.746 19.075
	24.860 19.082
	24.949 19.093
	25.037 19.108
	25.125 19.127
	25.212 19.149
	25.298 19.175
	25.384 19.205
	25.469 19.238
	25.552 19.275
	25.634 19.315
	25.714 19.359
	25.792 19.405
	25.869 19.455
	25.943 19.508
	26.016 19.563
	26.085 19.622
	26.152 19.683
	26.217 19.748
	26.278 19.814
	26.337 19.884
	26.392 19.956
	26.443 20.030
	26.491 20.107
	26.535 20.186
	26.575 20.267
	26.611 20.350
	26.643 20.436
	26.670 20.523
	26.693 20.612
	26.711 20.703
	26.724 20.796
	26.731 20.891
	26.734 20.987
	26.730 21.084
	26.722 21.179
	26.708 21.272
	26.689 21.364
	26.666 21.454
	26.637 21.541
	26.605 21.627
	26.567 21.711
	26.526 21.792
	26.481 21.871
	26.432 21.948
	26.379 22.023
	26.323 22.095
	26.263 22.164
	26.201 22.231
	26.135 22.295
	26.067 22.356
	25.996 22.415
	25.922 22.470
	25.847 22.523
	25.769 22.572
	25.690 22.618
	25.608 22.661
	25.526 22.701
	25.441 22.738
	25.356 22.771
	25.270 22.800
	25.182 22.826
	25.095 22.848
	25.006 22.866
	24.917 22.881
	24.829 22.892
	24.715 22.898
	24.603 22.895
	24.494 22.882
	24.385 22.859
	24.276 22.825
	24.167 22.779
	24.057 22.721
	23.944 22.651
	23.829 22.568
	23.711 22.472
	23.589 22.363
	23.526 22.302
	23.462 22.238
	23.396 22.171
	23.329 22.100
	23.261 22.024
	23.190 21.946
	23.119 21.863
	23.045 21.776
	22.969 21.685
	22.892 21.590
	/
%
%
\linethickness=1pt
\setplotsymbol ({\thinlinefont .})
%
%
\plot 24.373 14.791 25.608 14.791 /
\plot 25.608 14.791 26.844 14.791 /
%
%
\plot  8.023 15.191  7.547 14.715 /
\plot  7.547 15.172  8.023 14.696 /
%
%
\linethickness=0.500pt
\setplotsymbol ({\makebox(0,0)[l]{\tencirc\symbol{'175}}})
\put{\makebox(0,0)[l]{\circle*{ 0.318}}} at 24.373 14.791
\put{\makebox(0,0)[l]{\circle*{ 0.318}}} at 26.844 14.791
%
%
\linethickness=1pt
\setplotsymbol ({\thinlinefont .})
\ellipticalarc axes ratio  0.135:0.135  360 degrees 
	from 16.866 14.785 center at 16.730 14.785
%
%
\put{\SetFigFont{12}{14.4}{rm}(d)} [lB] at 32.163 16.796
\put{\SetFigFont{12}{14.4}{rm}(c)} [lB] at 22.670 16.828
\put{\SetFigFont{12}{14.4}{rm}(b)} [lB] at 12.446 16.923
\put{\SetFigFont{12}{14.4}{rm}(a)} [lB] at  3.842 16.828
\put{$= \fder_\alpha$} [lB] at  8.779 14.806
\put{$= 1$} [lB] at 17.503 14.630
\put{$=D{\bar\Delta}D$} [lB] at 27.783 14.581
\linethickness=0pt
\putrectangle corners at  2.581 23.764 and 36.597 14.554
\endpicture}

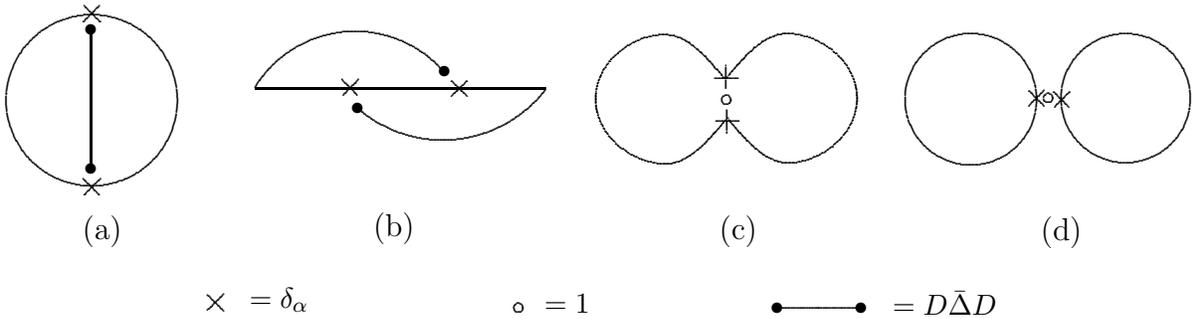
{\bf Figure 1:} The sewing diagrams. We call (a) the propagator 
insertion type, (b) the double folding type. The diagrams are made 
of the loops, the lines, and the black/white dots. See the text 
for details. 
\end{minipage}
\vspace{8mm}

Here we rather explain the method how to write down these 
necessary pieces for the pure Yang-Mills effective action, 
starting from graphical representations. 
These quantities \eq{iW1}-\eq{iW32} can be expressed by the 
graphical representations (sewing diagrams) in Figure 1(a)-(d). 
In the following, we present a general procedure 
to obtain the desired expressions in terms of our sewing 
technique. We shall follow the three steps explained below. 
(1) The first step: 
The basic parts to construct the sewing diagrams are the loop 
$\mbox{Trln}{\bar \Delta}^{-1}$, the line ${\bar\Delta}$, 
the white dot (identity propagator), and the cross 
$\delta_\alpha$. The edges of the ${\bar\Delta}$ line are 
expressed by the black dots (the covariant derivatives $D$'s), 
and the white dots themselves can formally be regarded as another 
kind of propagator edges as well. All these propagator edges 
should be joined with the $\delta_\alpha$ crosses, which are put 
on a loop or a line. The way of joining dots and crosses is that 
one has to connect a black dot with a single cross, and a white 
dot with a pair of two crosses. In this way, one can draw all 
possible sewing diagrams for a given topology of vacuum Feynman 
diagrams. In fact, {}~Figure 1(a)-(d) are all the possibilities 
to construct the two-loop vacuum topologies. 

(2) The second step:
After listing up the sewing diagrams, we then assign the 
integration variables $y_j$ for all crosses. It is enough to do 
this labeling just once, because they are just dummy variables. 
Then we perform the following identifications: 
If a white dot is attached to the crosses at $y_i$ and $y_j$, 
the identity propagator should be replaced by $\delta(y_i-y_j)$, 
and the pair of crosses becomes 
$( \delta/\delta\alpha^a_{\mu\nu}(y_i))^2$. 
If the edges of a line propagator are attached to the crosses at 
$y_i$ and $y_j$, then it becomes 
$(\delta_\alpha D)(y_i){\bar\Delta(y_i,y_j)}(D\delta_\alpha)(y_j)$. 
The color and Lorentz indices are to be read from the 
$J{\bar\Delta}J$ term in Eq.\eq{action} (taking account of 
the partial integrations):
\beq{*}
{\delta\over\delta\alpha^a_{\lambda\mu}} D^{ai}_\lambda
{\bar\Delta}^{ij}_{\mu\nu} D^{ej}_\rho
{\delta\over\delta\alpha^e_{\rho\nu}} \ .
\eeq 

(3) The third step:
{}~For a given sewing diagram $s_n$, which possesses $q$ crosses, 
$L$ loops, $k$ propagators, and $p$ white dots, 
we write down the following formal integral 
\beq
i\Gamma[s_n]=C_n \int\prod_{j=1}^q d^Dy_j\delta_\alpha(y_j) 
\Bigl(\prod^p \delta_\alpha 1 \delta_\alpha\Bigr) 
\Bigl(\prod^k \delta_\alpha D{\bar\Delta}D\delta_\alpha\Bigr)
\Bigl(\mbox{Trln}{\bar\Delta^{-1}}\Bigr)^L \Bigr|^{1PI}_{\alpha=0}\ ,
\eeq
and the numerical coefficient $C_n$ is determined 
by the following rules:
\beq{*}
\left.\begin{array}{ll}
i/2   &\quad\mbox{for}\quad   \fder_\alpha   \\
({i\over2})^n/n! &\quad\mbox{for $n$ propagators}\quad 
(D{\bar\Delta}D)^n  \\
({-i\over4})^n/n! &\quad\mbox{for $n$ identity propagators}\quad  \\
((-2)^L L!)^{-1} &\quad\mbox{for}\quad \mbox{$L$ loops}\quad 
(\mbox{Trln}{\bar\Delta}^{-1})^L\ .  \\
\end{array}\right. 
\eeq
Note that we may not attach an additional factor 2 in Fig. 1(a) 
(upside-down attachment of the $D{\bar\Delta}D$ propagator to the 
loop), since it is already included in the color summation, 
which can easily be understood from the color contraction in the 
second term on the r.h.s. of Eq.\eq{source}. Actually the cross 
term contributions from $J{\bar\Delta}J$ correspond to the factor 
2 of the $\phi^3$ case. In this way, this point superficially 
differs from the $\phi^3$ case~\cite{combi} 
(however is basically the same). 

Hereafter we deal with the Euclidean formulation for the 
world-line (path integral) representations. Applying the path 
integral representations of gluon loop and propagator 
(in terms of only the bosonic world-line field)~\cite{RSS} 
\beqa
\mbox{Trln}{\bar\Delta}^{-1}\dand=\dand
-\int_0^\infty{dS\over S}\oint{\cal D}x
\exp[-\int_0^S{1\over4}{\dot x}^2(\tau)d\tau](\mbox{Pexp}
\int_0^S\hskip-6pt{\bar M}[x(\tau)]d\tau)^{aa}_{\mu\mu}\ ,
\label{path1}\\
{\bar\Delta}^{ab}_{\mu\nu}(y_1,y_2)\dand=\dand 
\int_0^\infty\hskip-5pt d(\tau_2-\tau_1)
\int_{\scriptstyle x(\tau_2)=y_2 
        \atop\scriptstyle x(\tau_1)=y_1}
\hskip-3pt{\cal D}x
e^{-\int_{\tau_1}^{\tau_2}{1\over4}{\dot x}^2(\tau)d\tau}
(\mbox{Pexp}\hskip-3pt\int_{\tau_1}^{\tau_2}
\hskip-6pt{\bar M}[x(\tau)]d\tau)^{ab}_{\mu\nu}\label{path2}\\
\dand\equiv\dand\int_0^\infty\hskip-5pt d(\tau_2-\tau_1)\,
\int_{\scriptstyle x(\tau_2)=y_2 
        \atop\scriptstyle x(\tau_1)=y_1}{\cal D}x 
K^{ab}_{\mu\nu}(x|y_1,y_2;\tau_1,\tau_2)\ ,\label{pathK}
\eeqa
where 
\beq{*}
{\bar M}_{ab}[x(\tau)]=2i(F^c_{\mu\nu}+\alpha^c_{\mu\nu}
-\delta_{\mu\nu}{1\over2}A^c_\mu{\dot x}^\mu)\,(\lambda^c)_{ab}\ ,
\eeq
we derive the following path integral representations of the 
$\Gamma_i$ ($=\Gamma_1,\Gamma_2,\Gamma^{(1)}_3,\Gamma^{(2)}_3$) 
after some calculation (These will be derived in 
Section~\ref{sec3}):
\beqa
\Gamma_i&=& \gmnrs \int_0^\infty {dS\over S}\int_0^S d\tau_\beta 
\int_0^S d\tau_\alpha \int_0^\infty dT_3 \int d^Dy_1 d^Dy_2 \nn\\
&\times&\delta(y_1-x(\tau_\beta))\delta(y_2-x(\tau_\alpha))
{\cal V}_i(y_1,y'_1,y_2)
\int_{\scriptstyle x(S)=x(\tau_\beta) 
       \atop\scriptstyle x(0)=y'_1} [{\cal D}x]_S
\int_{\scriptstyle w(T_3)=y_2  \atop
\scriptstyle w(0)=y_1}[{\cal D}w]_{T_3}\Bigr|_{y'_1=y_1}\nn\\
&\times&[
(\mbox{P}e^{\int_{\tau_\beta}^{\tau_\alpha}M(x)})_{\delta\rho}
\lambda^e]^{km}
[(\mbox{P}e^{\int_{\tau_\alpha}^{\tau_\beta}M(x)})_{\epsilon\mu}
\lambda^a]^{gl}
(\mbox{P}e^{\int_0^{T_3}M(w)})^{nj}_{\gamma\sigma'}\ ,\label{EA}
\eeqa 
where $y'_1$ should be set to $y_1$ after ${\cal V}_i$ operating 
on the boundaries of path integrals, whose free parts are denoted by 
\beq{*}
[{\cal D}x]_S = {\cal D}x\,
\exp[-\int_0^S{1\over4}{\dot x}^2(\tau)d\tau] 
\quad\qquad\mbox{etc.}
\eeq 
Because of the anti-symmetric nature of $\alpha^a_{\mu\nu}$, we 
have introduced the symbol 
\beq{asymbol}
\delta^{\mu'\nu'}_{\mu\phantom{{}'}\nu} \equiv\, 
\delta^{\mu'}_{\mu} \delta^{\nu'}_{\nu}
-\delta^{\mu'}_{\nu} \delta^{\nu'}_{\mu}\ ,
\eeq 
and setting $\alpha^a_{\mu\nu}=0$, we have 
\beq{*}
M_{ab}(x)={\bar M}_{ab}[x(\tau)]\Bigr|_{\alpha=0}\ .
\eeq
The differential operators ${\cal V}_i$ are listed as follows:  
\beqa
{\cal V}_1&=&{1\over4}D^{ai}_{\mu'}(y_1) D^{ej}_{\rho'}(y_2)
\delta^{mg}\delta^{kl}\delta^{ni}
g_{\delta\nu}g_{\epsilon\sigma}g_{\gamma\nu'}\ ,\nn\\
{\cal V}_2&=&{1\over2}D^{ai}_{\mu'}(y'_1)D^{ej}_{\rho'}(y_2)
\delta^{mg}\delta^{ki}\delta^{ln}
g_{\delta\nu'}g_{\epsilon\sigma}g_{\gamma\nu}\ ,\label{vi}\\
{\cal V}_3^{(1)}&=&-{1\over8N}\delta^{ae}\delta^{ij}g_{\mu'\rho'}
\delta(y_1-y_2)\delta(T_3) \delta^{mg}\delta^{kl}\delta^{ni}
g_{\delta\nu}g_{\epsilon\sigma}g_{\gamma\nu'}\ ,\nn\\
{\cal V}_3^{(2)}&=&-{1\over16N}\delta^{ae}\delta^{ij}g_{\mu'\rho'}
\delta(y_1-y_2)\delta(T_3)
\delta^{mk}\delta^{gl}\delta^{ni}
g_{\delta\sigma}g_{\epsilon\nu}g_{\gamma\nu'}\ .\nn
\eeqa

Finally, some remarks are in order. 
(i) In deriving the above representation~\eq{EA}, we have employed the 
anti-path ordering in \eq{pathK}. Since the orderings of world-line 
paths are not related to the time orderings, this choice does not 
cause any trouble. Thus the direction of a matrix ordering and its 
proper time direction coincide in our case. One can easily transform 
them into the normal path ordering formalism by exchangig $y_1$ and $y_2$, 
or equivalently, $K^{ab}_{\mu\nu}(y_1,y_2)\ra K^{ba}_{\nu\mu}(y_1,y_2)$. 
(ii) We have artificially introduced $y'_i$ variables only for 
${\cal V}_2$ in order to keep track of the original covariant 
derivative positions, which shall be essential only for the 
analyses (Section~\ref{sec5}) of embeddings of 8-figure diagrams 
into $\Phi^3$ type diagrams. Since the embeddings are not a 
necessity but a matter of conveniences, one may put $y'_i=y_i$ as 
assumed in the above sewing rules. In the next section, we briefly 
view some details of how we obtain the expressions \eq{vi}. 
(iii) Remember that these are still intermediate results since 
non-world-line objects (i.e. covariant derivatives) still remain in 
the representations~\eq{vi}. However, at this level, we can see that 
the Lorentz and color structures in Eq.\eq{EA} with \eq{vi} coincide 
with those obtained by other methods in Appendix A. {}~For a reference, 
the graphical representations of the Lorentz and color indices for the 
${\cal V}_i$, $i=1,2$ are shown in Fig. 2. 
(iv) In order to see in detail the way how to obtain amplitudes in a 
Bern-Kosower form, one has to perform the substitution \eq{FT}, 
expanding the background field as mentioned before. 
The general procedures to do this are explained in 
Appendix B for the general action formula \eq{EA}.

\vspace{9mm}
\begin{minipage}[t]{14cm} 
%
%
\font\thinlinefont=cmr5
\begingroup\makeatletter\ifx\SetFigFont\undefined
\def\x#1#2#3#4#5#6#7\relax{\def\x{#1#2#3#4#5#6}}%
\expandafter\x\fmtname xxxxxx\relax \def\y{splain}%
\ifx\x\y   
\gdef\SetFigFont#1#2#3{%
  \ifnum #1<17\tiny\else \ifnum #1<20\small\else
  \ifnum #1<24\normalsize\else \ifnum #1<29\large\else
  \ifnum #1<34\Large\else \ifnum #1<41\LARGE\else
     \huge\fi\fi\fi\fi\fi\fi
  \csname #3\endcsname}%
\else
\gdef\SetFigFont#1#2#3{\begingroup
  \count@#1\relax \ifnum 25<\count@\count@25\fi
  \def\x{\endgroup\@setsize\SetFigFont{#2pt}}%
  \expandafter\x
    \csname \romannumeral\the\count@ pt\expandafter\endcsname
    \csname @\romannumeral\the\count@ pt\endcsname
  \csname #3\endcsname}%
\fi
\fi\endgroup
\mbox{\beginpicture
\setcoordinatesystem units <0.6900cm,0.6900cm>
\unitlength=0.6900cm
\linethickness=1pt
\setplotsymbol ({\makebox(0,0)[l]{\tencirc\symbol{'160}}})
\setshadesymbol ({\thinlinefont .})
\setlinear

%
%
\linethickness=14pt
\setplotsymbol ({\thinlinefont .})
\setdashes < 0.0877cm>
%
%
\plot  5.750 16.835  6.950 16.835 /
%
%
\plot  5.750 24.805  6.950 24.805 /
%
%
\plot 16.400 24.805 17.750 24.805 /
%
%
\plot 15.660 17.530 15.660 18.850 /
%
%
\setsolid
\circulararc 154.639 degrees from 16.192 24.130 center at 16.907 20.955
\circulararc 154.676 degrees from 17.780 17.780 center at 17.067 20.955
%
%
\plot  6.826 23.971  5.874 23.019 /
\plot  5.874 23.971  6.826 23.019 /
%
%
\plot  5.874 22.701  5.874 19.209 /
%
%
\plot  6.826 18.891  5.874 17.939 /
\plot  5.874 18.891  6.826 17.939 /
%
%
\circulararc 154.676 degrees from  7.144 17.780 center at  6.430 20.955
\circulararc 154.639 degrees from  5.556 24.130 center at  6.271 20.955
%
%
%
%
\plot  9.818 20.818  9.684 21.273  9.550 20.818 /
%
%
\plot  2.881 21.050  3.016 20.596  3.151 21.050 /
%
%
\plot  6.008 20.701  5.874 21.155  5.739 20.701 /
%
%
%
\plot 17.462 23.971 16.510 23.019 /
\plot 16.510 23.971 17.462 23.019 /
\plot 17.402 18.891 16.450 17.939 /
\plot 16.450 18.891 17.402 17.939 /
%
%
\plot 16.510 22.701 16.510 19.209 /
%
%
%
\plot 13.518 21.050 13.652 20.596 13.787 21.050 /
\plot 16.645 20.860 16.510 21.314 16.375 20.860 /
\plot 20.455 20.860 20.320 21.314 20.185 20.860 /
%
%
%
\put{$\nu$} [lB] at  6.985 17.350
\put{$\rho'$} [lB] at  7.144 22.860
\put{$\mu$} [lB] at  5.556 17.350
\put{$\mu'$} [lB] at  6.985 18.850
\put{$\nu'$} [lB] at  5.339 18.795
\put{$\sigma'$} [lB] at  5.397 22.860
\put{$\rho$} [lB] at  7.000 24.280
\put{$\sigma$} [lB] at  5.414 24.268
\put{$l$} [lB] at  5.556 16.680
\put{$k$} [lB] at  6.985 16.680
\put{$i$} [lB] at  5.397 19.495
\put{$j$} [lB] at  5.397 22.220
\put{$e$} [lB] at  7.144 22.220
\put{$a$} [lB] at  7.000 19.495
\put{$m$} [lB] at  7.000 24.730
\put{$g$} [lB] at  5.414 24.750
\put{$\nu$} [lB] at 16.100 18.891
\put{$\mu$} [lB] at 16.192 17.304
\put{$\sigma$} [lB] at 16.034 24.289
\put{$\sigma'$} [lB] at 16.034 22.860
\put{$\rho$} [lB] at 17.780 24.289
\put{$\rho'$} [lB] at 17.780 22.860
\put{$\mu'$} [lB] at 17.500 18.891
\put{$\nu'$} [lB] at 17.462 17.304
\put{$e$} [lB] at 17.780 22.220
\put{$a$} [lB] at 18.170 18.891
\put{$l$} [lB] at 15.695 17.259
\put{$k=i$} [lB] at 18.170 17.304
\put{$m$} [lB] at 17.780 24.730
\put{$g$} [lB] at 16.034 24.750
\put{$i$} [lB] at 15.450 18.891
\put{$j$} [lB] at 16.034 22.220
\put{\SetFigFont{12}{14.4}{rm}(b)} [lB] at 16.510 15.500
\put{\SetFigFont{12}{14.4}{rm}(a)} [lB] at  5.715 15.500
\linethickness=0pt
\putrectangle corners at  1.991 24.835 and 17.845 14.846
\endpicture}

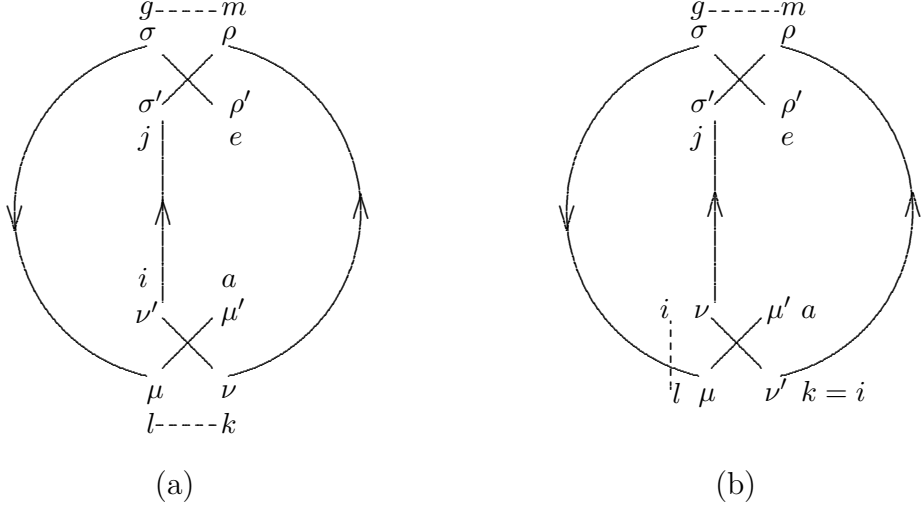
{\bf Figure 2:} The graphical representations of (a) ${\cal V}_1$ 
and (b) ${\cal V}_2$. The cross lines stand for the contractions 
defined by Eq.\eq{asymbol}. The dashed lines express the color 
indices contractions. 
\end{minipage}
\vspace{9mm}

\section{ Derivation of the $\Gamma_i$}\label{sec3}
\setcounter{section}{3}
\setcounter{equation}{0}
\indent

In this section, we explain how we further proceed with the 
computations of Eqs.\eq{iW1}-\eq{iW32}, in particular how to perform 
the extraction of 1PI parts. Basically, a second derivative of $K$ 
(defined in Eq.\eq{pathK}) gives rise to two kinds of terms composed 
of triple $K$ products due to the successive applications of the 
following formula: 
\beq{cut}
{\fder\over\fder\alpha^a_{\mu\nu}(y)}
K^{ij}_{\rho\sigma}(x|y_1,y_2;\tau_1,\tau_2)   
= \int_{\tau_1}^{\tau_2} d\tau \,
K^{ib}_{\rho\gamma}(x|y_1,y;\tau_1,\tau)
\left({\fder\over\fder\alpha^a_{\mu\nu}(y)}{\bar M}[x(\tau)]
\right)^{bc}_{\gamma\delta}
K^{cj}_{\delta\sigma}(x|y,y_2;\tau,\tau_2)\ .
\eeq
In $\Gamma_2$, one of these two kinds corresponds to the desired 
1PI parts, and the other kind to the one particle reducible (1PR) 
parts. On the other hand in $\Gamma_1$, both kinds contribute to 
the 1PI parts. We show the computational details of these facts 
case by case in the following subsections. These may rather sound 
like meticulous technical details, however as we shall see later, 
identifying 1PR parts is certainly useful in order to investigate 
the way how shrinking propagator limits form the $\Phi^4$ type 
``eight-figure'' diagrams $\Gamma^{(i)}_3$, $i=1,2$. {}~For example, 
in Section~\ref{sec3c}, we point out interesting relations between 
$\Gamma^{(1)}_3$ and $\Gamma_1$ as well as between $\Gamma^{(2)}_3$ 
and $\Gamma_1^R$, which is a 1PR type of $\Gamma_1$ in a sense. 
The $\Gamma_1$ calculation is also useful for comparing with the 
ghost loop in Section~\ref{sec7}.
\subsection{ The double folding types 
$\Gamma_2$ and $\Gamma^R_2$}\label{sec3a}
\indent

Let us start with the most non-trivial part, the separation of 
1PR contribution in $\Gamma_2$. Originally the double folding 
$\Gamma_2$ is a part of the following quantity 
(c.f. Eq.\eq{iW2} and Figure 1(b)):
\beq{*}
z_2 \equiv -{1\over8}\int d^Dy_1 d^Dy_2 D\fder_{\alpha}
{\bar\Delta}D\fder_{\alpha}\Bigr|_{\alpha=0}
= \Gamma_2+\Gamma^R_2 \ .
\eeq 
Let us explain the method how to separate the 1PI part from $z_2$. 
Operating a differentiation $\fder_\alpha$ on a propagator 
${\bar\Delta}$ ($\sim K$ of length $L$), two segments of 
'propagator' $K$ are created: 
\beqa
&&{\fder\over\fder\alpha^a_{\mu'\nu'}(y_1)}\int_0^\infty dL
K^{ij}_{\nu'\sigma'}(x|y_1,y_2;0,L)      \\
&=&\int_0^\infty d\tau_\beta \int_0^\infty dT_3\,
K^{ib}_{\nu'\mu}(x|y_1,y_1;0,\tau_\beta)
[2i\delta^{\mu'\nu'}_{\mu\phantom{{}'}\nu}(\lambda^a)_{bc}
\delta(y_1-x(\tau_\beta))]
K^{cj}_{\nu\sigma'}(w|y_1,y_2;0,T_3) \ ,\nn
\eeqa
where $L$ is shifted by the relation $T_3=L-\tau_\beta$, and 
$w$ is a redefined field (see Eq.\eq{wdef}). The 
graphical representation of this formula is shown in Fig. 3(a). 
A second differentiation applies to each of these two propagators by 
the Leibniz rule. One is (un-shifted integral) 
\beqa
&&{\fder\over\fder\alpha^e_{\rho'\sigma'}(y_2)}\int_0^\infty 
d\tau_\beta K^{ib}_{\nu'\mu}
(x|y_1,y_1;0,\tau_\beta)  \label{first}  \\
&&=\int_0^\infty d\tau_\beta \int_0^{\tau_\beta}d\tau_\alpha\,
K^{if}_{\nu'\rho}(x|y_1,y_2;0,\tau_\alpha)
[2i\delta^{\rho'\sigma'}_{\rho\phantom{{}'}\sigma}(\lambda^e)_{fg}
\delta(y_2-x(\tau_\alpha))]
K^{gb}_{\sigma\mu}(x|y_2,y_1;\tau_\alpha,\tau_\beta)\ , \nn
\eeqa
and the other is (with shifting $T=T_3 -T'_3$)
\beqa
&&{\fder\over\fder\alpha^e_{\rho'\sigma'}(y_2)}\int_0^\infty dT_3
K^{cj}_{\nu\sigma'}(w|y_1,y_2;0,T_3)  \label{second}  \\
&&=\int_0^\infty dT'_3 \int_0^\infty dT \,
K^{cf}_{\nu\rho}(w|y_1,y_2;0,T'_3)
[2i\delta^{\rho'\sigma'}_{\rho\phantom{{}'}\sigma}(\lambda^e)_{fg}
\delta(y_2-w(0))] K^{gj}_{\sigma\sigma'}(z|y_2,y_2;0,T) \nn\ .
\eeqa
The first one \eq{first} corresponds to the 1PI diagram $\Gamma_2$, 
and the latter one \eq{second} to the 1PR diagram $\Gamma^R_2$. 
Their graphical representations are shown in Fig. 3. 
Having separately three propagators, we have introduced the 
following splittings of the world-line field: 
\beq{wdef}
w(\tau)=x(\tau_\beta+\tau)\ , \qquad (0\leq\tau\leq T_3) 
\qquad\mbox{\rm{for 1PI}},
\eeq
and for the 1PR case 
\beq{*}
\left\{ \begin{array}{ll}
z(\tau)=x(\tau_\alpha+\tau),&\qquad 
(0\leq\tau\leq T=L-\tau_\alpha)\\ 
w(\tau)=x(\tau_\beta+\tau),&\qquad (0\leq\tau\leq T'_3=
\tau_\alpha-\tau_\beta)\ ,  
\end{array}\right.
\eeq
where the shifted length parameters $T_3$, $T'_3$ and $T$ all run 
from zero to infinity.

\vspace{8mm}
\begin{minipage}[t]{15cm} 
%
%
\font\thinlinefont=cmr5
\begingroup\makeatletter\ifx\SetFigFont\undefined
\def\x#1#2#3#4#5#6#7\relax{\def\x{#1#2#3#4#5#6}}%
\expandafter\x\fmtname xxxxxx\relax \def\y{splain}%
\ifx\x\y   
\gdef\SetFigFont#1#2#3{%
  \ifnum #1<17\tiny\else \ifnum #1<20\small\else
  \ifnum #1<24\normalsize\else \ifnum #1<29\large\else
  \ifnum #1<34\Large\else \ifnum #1<41\LARGE\else
     \huge\fi\fi\fi\fi\fi\fi
  \csname #3\endcsname}%
\else
\gdef\SetFigFont#1#2#3{\begingroup
  \count@#1\relax \ifnum 25<\count@\count@25\fi
  \def\x{\endgroup\@setsize\SetFigFont{#2pt}}%
  \expandafter\x
    \csname \romannumeral\the\count@ pt\expandafter\endcsname
    \csname @\romannumeral\the\count@ pt\endcsname
  \csname #3\endcsname}%
\fi
\fi\endgroup
\mbox{\beginpicture
\setcoordinatesystem units <0.73000cm,0.73000cm>
\unitlength=0.73000cm
\linethickness=1pt
\setplotsymbol ({\makebox(0,0)[l]{\tencirc\symbol{'160}}})
\setshadesymbol ({\thinlinefont .})
\setlinear
%
%
\linethickness=0.500pt
\setplotsymbol ({\thinlinefont .})
\circulararc 131.849 degrees from  5.874 22.860 center at  4.349 21.798
%
%
\linethickness=0.500pt
\setplotsymbol ({\thinlinefont .})
\circulararc 149.654 degrees from 22.225 22.225 center at 20.710 22.155
\circulararc 136.688 degrees from 17.145 22.860 center at 15.684 21.912
%
%
\linethickness=0.500pt
\setplotsymbol ({\makebox(0,0)[l]{\tencirc\symbol{'175}}})
\put{\makebox(0,0)[l]{\circle*{ 0.318}}} at 17.145 22.860
\put{\makebox(0,0)[l]{\circle*{ 0.318}}} at 19.367 22.860
%
%
\linethickness=0.500pt
\setplotsymbol ({\makebox(0,0)[l]{\tencirc\symbol{'175}}})
\put{\makebox(0,0)[l]{\circle*{ 0.318}}} at 10.160 22.225
%
%
\linethickness=0.500pt
\setplotsymbol ({\makebox(0,0)[l]{\tencirc\symbol{'175}}})
\put{\makebox(0,0)[l]{\circle*{ 0.318}}} at  5.874 22.860
%
%
\linethickness=0.500pt
\setplotsymbol ({\makebox(0,0)[l]{\tencirc\symbol{'175}}})
\putrule from  2.540 22.225 to 10.160 22.225
%
%
\linethickness=0.500pt
\setplotsymbol ({\makebox(0,0)[l]{\tencirc\symbol{'175}}})
\putrule from 13.970 22.225 to 22.225 22.225
%
%
\linethickness=1pt
\setplotsymbol ({\thinlinefont .})
\plot 16.986 22.001 17.435 22.449 /
\plot 16.986 22.449 17.435 22.001 /
%
%
\linethickness=1pt
\setplotsymbol ({\thinlinefont .})
\plot 19.209 22.001 19.657 22.449 /
\plot 19.209 22.449 19.657 22.001 /
%
%
\linethickness=1pt
\setplotsymbol ({\thinlinefont .})
\plot  5.874 22.001  6.322 22.449 /
\plot  5.901 22.449  6.350 22.001 /
%
%
\put{$w(T_3)$} [lB] at 10.160 22.860
\put{$w(0)$} [lB] at  6.509 22.860
\put{$x(\tau)$} [lB] at 15.558 24.130
\put{$z(\tau)$} [lB] at 20.796 24.130
\put{$w(\tau)$} [lB] at 17.900 22.543
\put{$T'_3$} [lB] at 18.256 21.273
\put{$\tau_\alpha$} [lB] at 19.526 21.273
\put{$\tau_\beta$} [lB] at 17.145 21.273
\put{$x(\tau_\beta)$} [lB] at  4.921 21.273
\put{$x(0)$} [lB] at  4.504 22.760
\put{\SetFigFont{12}{14.4}{rm}(a)} [lB] at  5.715 19.844
\put{\SetFigFont{12}{14.4}{rm}(b)} [lB] at 18.098 19.844
\linethickness=0pt
\putrectangle corners at  3.6150 24.282 and 22.250 19.768
\endpicture}

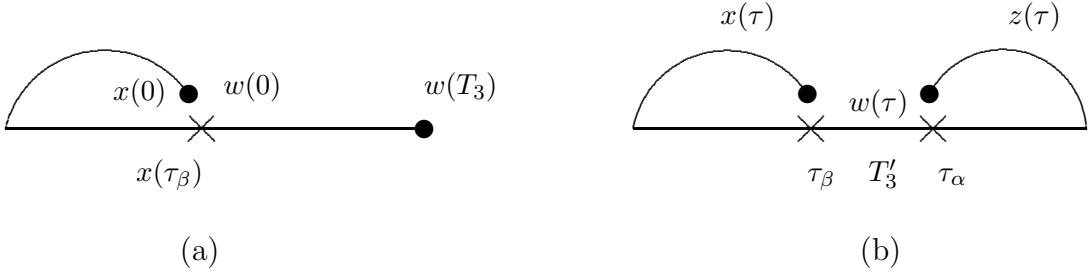
{\bf Figure 3:} The sewing procedures for $\Gamma_2$ and 
$\Gamma^R_2$. The 1PI diagram is obtained if $\tau_\alpha$ is 
inserted in the region $0<\tau_\alpha<\tau_\beta$ in the 
diagram~(a). Otherwise the 1PR diagram follows from the case~(b). 
\end{minipage}
\vspace{8mm}

The final results of the $\alpha$-differentiations are therefore 
\beqa
\Gamma_2 &=& {1\over2}\gmnrs \int_0^\infty d\tau_\beta
\int_0^\infty dT_3\int_0^{\tau_\beta} d\tau_\alpha\,\int d^Dy' \nn\\
&\times&D^{ai}_{\mu'}(y')D^{ej}_{\rho'}(x(\tau_\alpha)) 
\int_{\scriptstyle x(\tau_\beta)=y 
        \atop\scriptstyle x(0)=y'}[{\cal D}x]_{\tau_\beta}
\int_{\scriptstyle w(T_3)=x(\tau_\alpha) 
        \atop\scriptstyle w(0)=x(\tau_\beta)}[{\cal D}w]_{T_3}
\,\Bigr|_{y'=y}  \nn \\
&\times&[(\mbox{P}e^{\int_0^{\tau_\alpha}M(x)})_{\nu'\rho}\lambda^e
(\mbox{P}e^{\int_{\tau_\alpha}^{\tau_\beta}M(x)})_{\sigma\mu}
\lambda^a
(\mbox{P}e^{\int_0^{T_3}M(w)})_{\nu\sigma'}]^{ij}\ ,\label{w2}
\eeqa
\beqa
\Gamma_2^{R}&=&{1\over2}\gmnrs \int_0^\infty d\tau_\beta
\int_0^\infty dT'_3\int_0^\infty dT\,\int d^Dy'_1d^Dy'_2 \nn\\
&\times&D^{ai}_{\mu'}(y'_1) D^{ej}_{\rho'}(y'_2)
\int_{\scriptstyle x(\tau_\beta)=y_1 
        \atop\scriptstyle x(0)=y'_1}[{\cal D}x]_{\tau_\beta}
\int_{\scriptstyle z(T)=y'_2 
        \atop\scriptstyle z(0)=y_2}[{\cal D}z]_T
\int_{\scriptstyle w(T'_3)=z(0) 
        \atop\scriptstyle w(0)=x(\tau_\beta)}[{\cal D}w]_{T'_3}
\,\Bigr|_{y'_i=y_i}    \nn \\
&\times&[(\mbox{P}e^{\int_0^{\tau_\beta}M(x)})_{\nu'\mu}\lambda^a
(\mbox{P}e^{\int_0^{T'_3}M(w)})_{\nu\rho}\lambda^e
(\mbox{P}e^{\int_0^{T}M(z)})_{\sigma\sigma'}]^{ij}\ .\label{w2r}
\eeqa
{}~Fixing $\tau_\beta$ to $S$ in Eq.\eq{EA}, thereby getting a 
factor $S$ from the $\tau_\beta$ integral~\footnote{
This is because of the manifest rotational invariance in Eq.\eq{EA}. 
We call this procedure {\it fixing} hereafter.}, 
one can see that the above result \eq{w2} coincides with 
the one given in Eq.\eq{EA} with ${\cal V}_2$. 

\subsection{ The propagator insertion types 
$\Gamma_1$ and $\Gamma_1^R$ }\label{sec3b}
\indent

{}~First let us consider the $\Gamma_1$ case. One can formally express 
the path integrand of \eq{path1} in terms of $K$ defined \eq{pathK}; 
that is nothing but $K^{aa}_{\mu\mu}(x|x,x;0,S)$ with imposing the 
condition $x(0)=x(S)$ (world-line length $S$). Note that this is 
definitely different from the $K^{aa}_{\mu\mu}(x|x,x,\tau,\tau)$ 
which are living on a point (world-line length zero). Now, 
the Leibniz rule on a second derivative of the $K$ creates the 
following three segments of the propagator:
\beqa
&&{\fder\over\fder\alpha^a_{\mu'\nu'}(y_1)}
{\fder\over\fder\alpha^e_{\rho'\sigma'}(y_2)}
\int_0^\infty{dS\over S}K^{ii}_{\alpha\alpha}(x|x,x;0,S) \nn\\
&&=\int_0^\infty{dS\over S}\int_0^S d\tau_\beta\int_0^S 
d\tau_\alpha \,(2i)^2 \delta(y_1-x(\tau_\beta))
\delta(y_2-x(\tau_\alpha))\gmnrs \nn\\
&&\times\Bigl\{ \theta(\tau_\beta-\tau_\alpha)
[(\mbox{P}e^{\int_0^{\tau_\alpha}M(x)})_{\alpha\rho}\lambda^e
(\mbox{P}e^{\int_{\tau_\alpha}^{\tau_\beta}M(x)})_{\sigma\mu}
\lambda^a
(\mbox{P}e^{\int_{\tau_\beta}^SM(x)})_{\nu\alpha}]^{ii}\nn\\
&&+ \theta(\tau_\alpha-\tau_\beta)
[(\mbox{P}e^{\int_0^{\tau_\beta}M(x)})_{\alpha\mu}\lambda^a
(\mbox{P}e^{\int_{\tau_\beta}^{\tau_\alpha}M(x)})_{\nu\rho}\lambda^e
(\mbox{P}e^{\int_{\tau_\alpha}^SM(x)})_{\sigma\alpha}]^{ii}\Bigr\}\ .
\label{aaK}
\eeqa
Here we have two terms with step functions, and the first 
step-function term 
becomes 
\beq{*}
\theta(\tau_\beta-\tau_\alpha) 
[\,(\mbox{P}e^{\int_{\tau_\beta}^{\tau_\alpha}M(x)})_{\nu\rho}
\lambda^e (\mbox{P}e^{\int_{\tau_\alpha}^{\tau_\beta}M(x)})
_{\sigma\mu}\lambda^a\,]^{ii} \ ,
\eeq
owing to the formula 
\beq{*}
(\mbox{P}e^{\int_{\tau_\beta}^SM(x)})_{\nu\alpha}
(\mbox{P}e^{\int_0^{\tau_\alpha}M(x)})_{\alpha\rho}
=(\mbox{P}' e^{\int_{\tau_\beta}^{\tau_\alpha}M(x)})_{\nu\rho}\ ,
\eeq
where $\mbox{P}'$ follows the path from $\tau_\beta$ to 
$\tau_\alpha$ via $S$. The second step-function term in Eq.\eq{aaK} 
amounts to the same quantity, and we hence drop the step functions 
because of the property $\theta(x)+\theta(-x)=1$. Therefore we arrive 
at Eq.\eq{EA} with ${\cal V}_1$ given in Eq.\eq{vi}:
\beqa
\Gamma_1&=&{1\over4}\gmnrs
\int_0^\infty {dS\over S}\int_0^S d\tau_\beta\int_0^S d\tau_\alpha
\int_0^\infty dT_3 \oint[{\cal D}x]_S \nn\\
&\times&D^{ai}_{\mu'}(x(\tau_\beta))
D^{ej}_{\rho'}(x(\tau_\alpha))
\int_{\scriptstyle w(T_3)=x(\tau_\alpha) 
        \atop\scriptstyle w(0)=x(\tau_\beta)}[{\cal D}w]_{T_3}\nn\\
&\times&\mbox{Tr}[
(\mbox{P}' e^{\int_{\tau_\beta}^{\tau_\alpha}M(x)})_{\nu\rho}
\lambda^e
(\mbox{P}e^{\int_{\tau_\alpha}^{\tau_\beta}M(x)})_{\sigma\mu}
\lambda^a]
(\mbox{P}e^{\int_0^{T_3}M(w)})^{ij}_{\nu'\sigma'}\ .\label{w1}
\eeqa  
{}~For later convenience, we also write down the  
1PR part made of two loops and one propagator (see Fig. 4(a)). 
According to the rules explained in Section~\ref{sec2}, this can be 
extracted from the quantity \eq{pureac} as follows:
\beqa
\Gamma_1^R&\equiv&{1\over2}(-i)\int dy_1 dy_2 (-{1\over2}
\mbox{Trln}{\bar\Delta}^{-1})^2({i\over2})^3\fder_\alpha 
D{\bar\Delta} \fder_\alpha D\Bigr|^{connected}_{\alpha=0} \label{G1Rdef}\\
&=& {1\over8}\gmnrs\int_0^\infty {dS\over S}\int_0^\infty{dT\over T}
\int_0^\infty dT_3 \int_0^S d\tau_\beta \int_0^T d\tau_\alpha\,
\oint[{\cal D}x]_S\oint[{\cal D}z]_T \nn\\
&\times&D^{ai}_{\mu'}(x(\tau_\beta))
D^{ej}_{\rho'}(z(\tau_\alpha))
\int_{\scriptstyle w(T_3)=z(\tau_\alpha) 
\atop\scriptstyle w(0)=x(\tau_\beta)}[{\cal D}w]_{T_3}   \nn\\
&\times&\mbox{Tr}[(\mbox{P}e^{\int_0^SM(x)})_{\nu\mu}\lambda^a]
\mbox{Tr}[(\mbox{P}e^{\int_0^TM(z)})_{\sigma\rho}\lambda^e]
(\mbox{P}e^{\int_0^{T_3}M(w)})_{\nu'\sigma'}^{ij}\ .\label{w1r}
\eeqa

\vspace{8mm}
\begin{minipage}[t]{14.5cm}
%
%
\font\thinlinefont=cmr5
\begingroup\makeatletter\ifx\SetFigFont\undefined
\def\x#1#2#3#4#5#6#7\relax{\def\x{#1#2#3#4#5#6}}%
\expandafter\x\fmtname xxxxxx\relax \def\y{splain}%
\ifx\x\y   
\gdef\SetFigFont#1#2#3{%
  \ifnum #1<17\tiny\else \ifnum #1<20\small\else
  \ifnum #1<24\normalsize\else \ifnum #1<29\large\else
  \ifnum #1<34\Large\else \ifnum #1<41\LARGE\else
     \huge\fi\fi\fi\fi\fi\fi
  \csname #3\endcsname}%
\else
\gdef\SetFigFont#1#2#3{\begingroup
  \count@#1\relax \ifnum 25<\count@\count@25\fi
  \def\x{\endgroup\@setsize\SetFigFont{#2pt}}%
  \expandafter\x
    \csname \romannumeral\the\count@ pt\expandafter\endcsname
    \csname @\romannumeral\the\count@ pt\endcsname
  \csname #3\endcsname}%
\fi
\fi\endgroup
\mbox{\beginpicture
\setcoordinatesystem units <0.70000cm,0.70000cm>
\unitlength=0.70000cm
\linethickness=1pt
\setplotsymbol ({\makebox(0,0)[l]{\tencirc\symbol{'160}}})
\setshadesymbol ({\thinlinefont .})
\setlinear
%
%
\linethickness=14pt
\setplotsymbol ({\thinlinefont .})
\circulararc 136.688 degrees from 15.875 22.860 center at 14.414 21.912
%
%
\linethickness=14pt
\setplotsymbol ({\makebox(0,0)[l]{\tencirc\symbol{'175}}})
\put{\makebox(0,0)[l]{\circle*{ 0.318}}} at  4.763 22.225
%
%
\linethickness=14pt
\setplotsymbol ({\makebox(0,0)[l]{\tencirc\symbol{'175}}})
\put{\makebox(0,0)[l]{\circle*{ 0.318}}} at  7.620 22.225
%
%
\linethickness=14pt
\setplotsymbol ({\thinlinefont .})
\ellipticalarc axes ratio  1.308:1.308  360 degrees 
	from 10.823 22.225 center at  9.514 22.225
%
%
\linethickness=14pt
\setplotsymbol ({\thinlinefont .})
\ellipticalarc axes ratio  1.308:1.308  360 degrees 
	from  4.180 22.225 center at  2.872 22.225
%
%
\linethickness=14pt
\setplotsymbol ({\makebox(0,0)[l]{\tencirc\symbol{'175}}})
\put{\makebox(0,0)[l]{\circle*{ 0.318}}} at 15.875 22.860
%
%
\linethickness=14pt
\setplotsymbol ({\makebox(0,0)[l]{\tencirc\symbol{'175}}})
\put{\makebox(0,0)[l]{\circle*{ 0.318}}} at 17.939 22.225
%
%
\linethickness=14pt
\setplotsymbol ({\thinlinefont .})
\ellipticalarc axes ratio  1.308:1.308  360 degrees 
	from 21.114 22.263 center at 19.806 22.263
%
%
\linethickness=29pt
\setplotsymbol ({\thinlinefont .})
\plot  3.969 22.001  4.417 22.449 /
\plot  3.969 22.449  4.417 22.001 /
%
%
\linethickness=0.500pt
\setplotsymbol ({\makebox(0,0)[l]{\tencirc\symbol{'175}}})
\putrule from  4.868 22.225 to  7.620 22.225
%
%
\linethickness=29pt
\setplotsymbol ({\thinlinefont .})
\plot  7.965 22.449  8.414 22.001 /
\plot  7.965 22.001  8.414 22.449 /
%
%
\linethickness=29pt
\setplotsymbol ({\thinlinefont .})
\plot 15.903 22.449 16.351 22.001 /
\plot 15.875 22.001 16.324 22.449 /
%
%
\linethickness=0.500pt
\setplotsymbol ({\makebox(0,0)[l]{\tencirc\symbol{'175}}})
\putrule from 12.700 22.225 to 17.964 22.225
%
%
\linethickness=29pt
\setplotsymbol ({\thinlinefont .})
\plot 18.256 22.001 18.705 22.449 /
\plot 18.284 22.449 18.733 22.001 /
%
%
\put{\SetFigFont{12}{14.4}{rm}(b)} [lB] at 16.034 20.003
\put{\SetFigFont{12}{14.4}{rm}(a)} [lB] at  5.874 20.003
\linethickness=0pt
\putrectangle corners at  1.347 23.675 and 21.131 19.926
\endpicture}

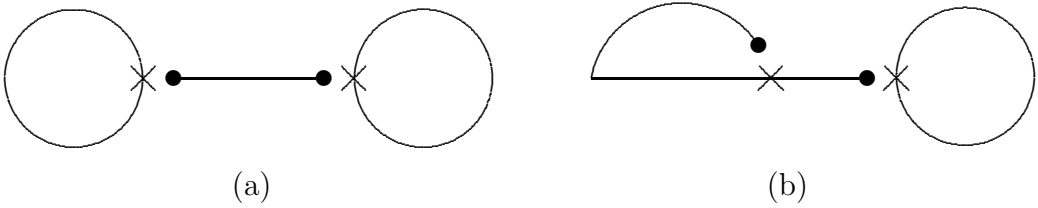
{\bf Figure 4:} (a) The sewing diagram for $\Gamma^R_1$. 
(b) Another 1PR sewing diagram which we do not consider.  
\end{minipage}
\vspace{8mm}

\subsection{The shrinkage types $\Gamma^{(1)}_3$ and 
$\Gamma^{(2)}_3$}\label{sec3c}
\indent


Comparing the defining equations of $\Gamma_1$, $\Gamma^R_1$ and 
$\Gamma^{(i)}_3$ (q.v. Eqs.\eq{iW1},\eq{iW31},\eq{iW32} and 
\eq{G1Rdef}), we notice that $\Gamma^{(1)}_3$ and $\Gamma^{(2)}_3$ 
can be obtained from $\Gamma_1$ and $\Gamma_1^R$ respectively 
in terms of the following replacement: 
\beq{replace}
\Bigl(D^{ai}_{\mu'}{\fder\over\fder\alpha^a_{\mu'\nu'}}\Bigr)
(y_1){\bar\Delta}^{ij}_{\nu'\sigma'}(y_1,y_2) \Bigl(D^{ej}_{\rho'}
{\fder\over\fder\alpha^e_{\rho'\sigma'}}\Bigr)(y_2)
\ra -{1\over2}\delta^{ae}g_{\mu'\rho'}g_{\nu'\sigma'}
\delta(y_1-y_2){\fder\over\fder\alpha^a_{\mu'\nu'}(y_1)}
{\fder\over\fder\alpha^e_{\rho'\sigma'}(y_2)}\ . \label{replace}
\eeq
In view of this fact, we have only to make the following 
replacement in Eqs.\eq{w1} and \eq{w1r}:
\beq{rep2}
D^{ai}_{\mu'}(x(\tau_\beta)) D^{ej}_{\rho'}(z(\tau_\alpha))
(\mbox{P}e^{\int_0^{T_3}M(w)})_{\nu'\sigma'}^{ij}
\ra -{1\over2}\delta^{ae}g_{\mu'\rho'}g_{\nu'\sigma'}
\delta(T_3)\delta(x(\tau_\beta)-x(\tau_\alpha))\ ,
\eeq
and we immediately obtain
\beqa
\Gamma_3^{(1)}&=&-{1\over8}\gmnrs 
\int_0^\infty {dS\over S}\int_0^S d\tau_\beta 
\int_0^S d\tau_\alpha \oint[{\cal D}x]_S \nn \\
&\times&g_{\mu'\rho'}g_{\nu'\sigma'}
\delta(x(\tau_\beta)-x(\tau_\alpha))
\mbox{Tr}[
(\mbox{P}'e^{\int_{\tau_\beta}^{\tau_\alpha}M(x)})_{\nu\rho}\lambda^a
(\mbox{P}e^{\int_{\tau_\alpha}^{\tau_\beta}M(x)})_{\sigma\mu}
\lambda^a]\ ,\label{W31}
\eeqa
\beqa
\Gamma_3^{(2)}&=&-{1\over16}\gmnrs 
\int_0^\infty {dS\over S}\int_0^\infty{dT\over T}
\int_0^S d\tau_\beta \int_0^T d\tau_\alpha\,
\oint[{\cal D}x]_S \oint[{\cal D}z]_T \nn \\
&\times&g_{\mu'\rho'}g_{\nu'\sigma'}
\delta(x(\tau_\beta)-z(\tau_\alpha))
\mbox{Tr}[(\mbox{P}e^{\int_0^SM(x)})_{\nu\mu}\lambda^a]
\mbox{Tr}[(\mbox{P}e^{\int_0^TM(z)})_{\sigma\rho}\lambda^a]\ .
\label{W32}
\eeqa
Eq.\eq{W31} coincides with the action \eq{EA} with 
${\cal V}^{(1)}_3$ given in Eq.\eq{vi}. The coincidence of 
Eq.\eq{W32} with \eq{EA} can be confirmed as follows. Fixing 
$\tau_\beta=S$ and $\tau_\alpha=0$, and shifting $T=U-S$, we see 
that the quantity \eq{W32} behaves as (up to the overall constant 
and metric symbols etc.) 
\beq{*}
\Gamma_3^{(2)}\sim\int_0^\infty dU\int_0^\infty dS\oint[{\cal D}x]_S
\oint[{\cal D}z]_{U-S}
\delta(x(S)-z(0))
(\mbox{P}e^{\int_0^SM(x)})^{ab}_{\nu\mu}
(\mbox{P}e^{\int_0^{U-S}M(z)})^{cd}_{\sigma\rho}\ .\nn
\eeq
We then merge the $z(\tau)$ integration into the $x(\tau)$ 
integration through the relation  
\beq{*}
z(\tau)=x(S+\tau)\ ;\quad \qquad 0\leq\tau\leq T, 
\eeq
thereby having 
\beq{g3sym}
\Gamma_3^{(2)}\sim\int_0^\infty dU\int_0^\infty dS\oint[{\cal D}x]_U
\delta(x(S)-x(U))
(\mbox{P}e^{\int_0^S M(x)})^{ab}_{\nu\mu}
(\mbox{P}e^{\int_S^U M(x)})^{cd}_{\sigma\rho}\ .\nn
\eeq
Renaming $S\ra U$ and $\tau_\beta\ra S$ after fixing 
$\tau_\alpha=S$ in Eq.\eq{EA}, we therefore prove that Eq.\eq{g3sym} 
produces Eq.\eq{EA} with ${\cal V}^{(2)}_3$ given in Eq.\eq{vi}. 

\section{The shrinking limit conjecture }\label{sec4}
\setcounter{section}{4}
\setcounter{equation}{0} 
\indent

In this section, we make a conjecture emerging from the analysis 
of Section~\ref{sec3c}. We have reproduced the shrinkage types 
$\Gamma_3^{(1)}$ and $\Gamma_3^{(2)}$ from $\Gamma_1$ and 
$\Gamma_1^R$, and we infer that this fact might also apply to the 
double folding types by taking an appropriate shrinking limit of a 
propagator with two covariant derivatives. 
 
Let us consider the replacement \eq{rep2}, in which $\delta^{ae}$ 
should be understood as $\delta^{ai}\delta^{ej}\delta^{ij}$. It may 
be expressed as the following limit operation $\delta_R$:
\beq{lim1}
\int[{\cal D}]_{T_i}
K^{ij}_{\rho\sigma}(x|y_1,y_2;0,T_i)\quad\tlim{T_i}\quad
2 \delta^{ij}g_{\rho\sigma}\ ,
\eeq
with 
\beq{lim2}
D^{ai}_\mu (y_1) D^{ej}_\nu (y_2) \quad\ra\quad 
-{1\over4}\delta^{ai}\delta^{ej}g_{\mu\nu}\delta(y_1-y_2)\ .
\eeq
The origins of the numerical factors will be clarified later 
in Section~\ref{sec5}. In this case, we seem to have 
\beqa
&&\delta_R(T_3):\quad\Gamma_1\quad\tlim{T_3}
\quad\Gamma_3^{(1)}\ ,\nn\\
&&\delta_R(T_3):\quad\Gamma_1^R\quad\tlim{T_3}
\quad\Gamma_3^{(2)}\ ,\nn\\  
&&\delta_R(T'_3):\quad\Gamma_2^R\quad\tlim{T'_3}
\quad\Gamma_3^{(2)}\ , 
\label{deltar}\\
&&\delta_R(T_3):\quad\Gamma_2\quad\tlim{T_3}
\quad-\Gamma_3^{(1)}\ .\nn
\eeqa  
However, there should be something wrong with the limit of 
$\Gamma_2$, since the sum of $\Gamma_1$ and $\Gamma_2$ vanishes 
in the limit \eq{deltar}. 

In fact, $\delta_R$ can not apply to $\Gamma_2$ and $\Gamma_2^R$, 
since the position of $D^{ai}$ is not exactly the same point 
as the starting point $y_1$ of the shrinking propagator, 
but rather the point $y'_1$ before setting $y'_1=y_1$ 
(q.v. Eqs.\eq{EA}, \eq{vi}, \eq{w2}, \eq{w2r}). The graphical 
situations in sewing diagrams are shown in Fig. 5. 
The limit $\delta_R$ is only relevant to the diagram Fig. 5(a). 
While in the diagram Fig. 5(b), we assume the following limit 
operation $\delta_0$: 
\beq{zero}
D^{ai}_\mu (y'_1) K^{kl}_{\rho\sigma}(y_1,y_2;0,T_i)D^{ej}_\nu (y_2) 
\quad\tlim{T_i}\quad -{1\over2}\delta^{ai}\delta^{ej}\delta^{kl}
g_{\mu\rho}g_{\sigma\nu} \delta(y'_1-y_2)\ ,
\eeq
which leads to 
\beqa
&&\delta_0(T_3):\quad\Gamma_2 \quad\tlim{T_3}\quad 0\ ,\nn\\
&&\delta_0(T'_3):\quad\Gamma_2^R\quad\tlim{T'_3}\quad 0\ .
\eeqa 

\vspace{8mm}
\begin{minipage}[t]{15cm} 
%
%
\font\thinlinefont=cmr5
\begingroup\makeatletter\ifx\SetFigFont\undefined
\def\x#1#2#3#4#5#6#7\relax{\def\x{#1#2#3#4#5#6}}%
\expandafter\x\fmtname xxxxxx\relax \def\y{splain}%
\ifx\x\y   
\gdef\SetFigFont#1#2#3{%
  \ifnum #1<17\tiny\else \ifnum #1<20\small\else
  \ifnum #1<24\normalsize\else \ifnum #1<29\large\else
  \ifnum #1<34\Large\else \ifnum #1<41\LARGE\else
     \huge\fi\fi\fi\fi\fi\fi
  \csname #3\endcsname}%
\else
\gdef\SetFigFont#1#2#3{\begingroup
  \count@#1\relax \ifnum 25<\count@\count@25\fi
  \def\x{\endgroup\@setsize\SetFigFont{#2pt}}%
  \expandafter\x
    \csname \romannumeral\the\count@ pt\expandafter\endcsname
    \csname @\romannumeral\the\count@ pt\endcsname
  \csname #3\endcsname}%
\fi
\fi\endgroup
\mbox{\beginpicture
\setcoordinatesystem units <0.70000cm,0.70000cm>
\unitlength=0.70000cm
\linethickness=1pt
\setplotsymbol ({\makebox(0,0)[l]{\tencirc\symbol{'160}}})
\setshadesymbol ({\thinlinefont .})
\setlinear
%
%
\linethickness=14pt
\setplotsymbol ({\makebox(0,0)[l]{\tencirc\symbol{'175}}})
\put{\makebox(0,0)[l]{\circle*{ 0.318}}} at  3.016 22.225
\put{\makebox(0,0)[l]{\circle*{ 0.318}}} at  9.684 22.225
\put{\makebox(0,0)[l]{\circle*{ 0.318}}} at 13.335 22.225
\put{\makebox(0,0)[l]{\circle*{ 0.318}}} at 20.796 22.225
%
%
\linethickness=29pt
\setplotsymbol ({\thinlinefont .})
\plot 15.903 22.449 16.351 22.001 /
\plot 15.875 22.001 16.324 22.449 /
\plot 18.256 22.001 18.705 22.449 /
\plot 18.284 22.449 18.733 22.001 /
%
%
\linethickness=0.500pt
\setplotsymbol ({\makebox(0,0)[l]{\tencirc\symbol{'175}}})
\putrule from  3.016 22.225 to  9.684 22.225
\putrule from 13.335 22.225 to 20.796 22.225
%
%
\put{\SetFigFont{12}{14.4}{rm}(a)} [lB] at  5.874 20.003
\put{\SetFigFont{12}{14.4}{rm}(b)} [lB] at 16.828 20.003
%
%
\put{$D_\mu(y_1)$} [lB] at  2.616 22.800
\put{$\rho$} [lB] at  3.351 21.431
\put{$\sigma$} [lB] at  8.814 21.431
\put{$D_\nu(y_2)$} [lB] at  9.207 22.800
\put{$D_\mu(y'_1)$} [lB] at 13.018 22.800
\put{$\rho$} [lB] at 18.398 21.273
\put{$y_1$} [lB] at 18.398 22.800
\put{$\sigma$} [lB] at 20.103 21.273
\put{$D_\nu(y_2)$} [lB] at 20.079 22.800
\linethickness=0pt
\putrectangle corners at  2.841 23.482 and 20.972 19.926
\endpicture}

\begin{center}
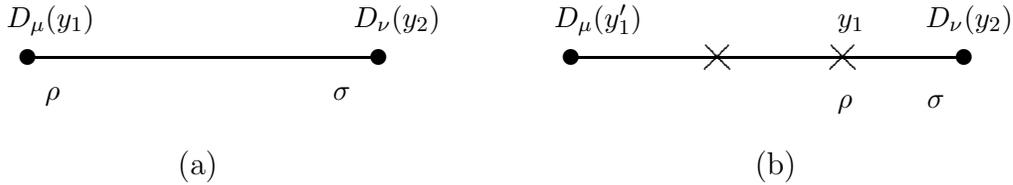
{\bf Figure 5:} The positions of shrinking points.  
\end{center}
\end{minipage}
\vspace{8mm}

{}~For later convenience, we here arrange the 1PR parts 
$\Gamma^R_i$; $i=1,2$, in a compact form similar to Eq.\eq{EA}. 
Fixing $\tau_\beta=S$ and $\tau_\alpha=0$ in Eq.\eq{w1r}, 
and renaming some integration variables in Eq.\eq{w2r}, 
we have 
\beqa
\Gamma_i^R&=& \gmnrs\int_0^\infty dS\int_0^\infty dT_3
\int_0^\infty dT  \nn \\
&\times& {\cal V}_i^{1PR}\,
\int_{\scriptstyle x(S)=y_1 \atop\scriptstyle x(0)=y'_1}
[{\cal D}x]_S\, 
\int_{\scriptstyle z(T)=y'_2 \atop\scriptstyle z(0)=y_2}
[{\cal D}z]_T\, 
\int_{\scriptstyle w(T_3)=z(0) \atop\scriptstyle w(0)=x(S)}
[{\cal D}w]_{T_3}\,\Bigr|_{y'_i=y_i} \nn \\
&\times&[(\mbox{P}e^{\int_0^SM(x)})_{\alpha\mu}\lambda^a]^{km}
(\mbox{P}e^{\int_0^{T_3}M(w)})_{\beta\gamma}^{nf}
[\lambda^e(\mbox{P}e^{\int_0^T M(z)})_{\sigma\delta}]^{gl} 
\label{EAR}
\eeqa
with 
\beqa
&&{\cal V}_1^{1PR}= {1\over8} 
           \delta^{km} \delta^{gl} \delta^{ni} \delta^{fj}
    g_{\alpha\nu} g_{\beta\nu'} g_{\gamma\sigma'}g_{\delta\rho}
D^{ai}_{\mu'}(x(S))D^{ej}_{\rho'}(z(0)) \ ,\\
&&{\cal V}_2^{1PR}= {1\over2} 
          \delta^{ki} \delta^{mn} \delta^{fg} \delta^{lj}
    g_{\alpha\nu'} g_{\beta\nu} g_{\gamma\rho} g_{\delta\sigma'}
D^{ai}_{\mu'}(y'_1)D^{ej}_{\rho'}(y'_2) \ ,
\eeqa
where $T_3$ is assigned to be the parameter (length of the 
propagator) which is taken to be zero in the limits $\delta_R$ 
and $\delta_0$. The conjecture is then written in the form:
\beqa
\delta_R \Gamma_1 + \delta_0 \Gamma_2 \quad 
\tlim{T_3}\quad \Gamma_3^{(1)}\ ,\label{shrink1}\\
\delta_R \Gamma_1^R+\delta_0 \Gamma_2^R \quad 
\tlim{T_3}\quad \Gamma_3^{(2)}\ .\label{shrink2}
\eeqa 
In the next section, we shall present a piece of evidence for 
the shrinking limits \eq{shrink1} and \eq{shrink2} in view of the 
full world-line representation, 
thus clarifying the origin of the factors attached in 
the rules \eq{lim1}, \eq{lim2} and \eq{zero}. 

\section{The pure Yang-Mills world-line formulae}\label{sec5}
\setcounter{section}{5}
\setcounter{equation}{0} 
\indent


In this section, we simplify all the previous results including 
the 1PR parts $\Gamma_i^R$. Actually we show that the quantities 
\eq{EA} and \eq{EAR} can be contained in a few concise expressions 
by analyzing the shrinking limits in the full world-line picture. 

Let us start with the following observation. If a covariant 
derivative is acting on an edge of the gauge particle propagator, 
the following formula holds ({\it on the propagator}):
\beq{dotform}
D^{ab}_\mu(x_1)
\int_{\scriptstyle w(T)=x_1 \atop\scriptstyle w(0)=x_2}
[{\cal D}w]_T=
{1\over2}\delta^{ab}
\int_{\scriptstyle w(T)=x_1 \atop\scriptstyle w(0)=x_2}
[{\cal D}w]_T\,
{\dot w}_\mu(T) \ .
\eeq
When one applies the above formula {\it twice}, one needs an extra 
minus sign on the r.h.s. of the formula; i.e. first getting the 
factor ${1\over2}$, and then $-{1\over2}$. This sign is 
consistent with the Minkowski case, which provides a factor 
$({i\over2})^2$ instead of the extra minus sign. Applying this and 
the following formulae to $\Gamma_2$ in Eq.\eq{EA} or in \eq{w2}:  
\beq{*}
(A \lambda^e B \lambda^a C)^{ae} 
=  -\mbox{Tr}[A \lambda^e B \lambda^a]C^{ae} \ ,
\eeq
with 
\beq{p}
\Bigl(\mbox{P}e^{\int_{\tau_\alpha}^{\tau_\beta}M(\tau)d\tau}
\Bigr)_{\nu\mu}^{ab}=
\Bigl(\mbox{P}e^{\int_{\tau_\alpha}^{\tau_\beta}M(-\tau
+\tau_\alpha+\tau_\beta)d\tau}\Bigr)_{\mu\nu}^{ba}\ ,
\eeq
one can see that $\Gamma_2$ consists of two 
parts, one of which has the same Lorentz index structure as $\Gamma_1$, 
and the other has a different one. We then transform $\Gamma_1+\Gamma_2$ 
into the following simplified combinations (classified by the 
types of Lorentz index structures).
\beq{1234}
\Gamma_1+\Gamma_2 = \Gamma_3 + \Gamma_4\ ,
\eeq
with
\beqa
\Gamma_3 
&=&\hskip-3pt -{1\over8}\int_0^\infty\hskip-3pt dS\hskip-3pt
\int_0^S\hskip-3pt d\tau_\alpha\hskip-3pt 
\int_0^\infty\hskip-4pt dT_3 \oint[{\cal D}x]_S\hskip-3pt
\int_{\scriptstyle w(T_3)=x(\tau_\alpha) 
       \atop\scriptstyle w(0)=x(0)}[{\cal D}w]_{T_3}\,
({\dot w}_\mu(0)-{\dot x}_\mu(0)){\dot w}_\rho(T_3)\nn \\
&\times& \mbox{Tr}[\, (\mbox{P}
e^{\int_0^{\tau_\alpha}M(x)})_{\nu[\rho}\lambda^e
(\mbox{P}e^{\int_{\tau_\alpha}^S M(x)})_{\sigma]\mu}
\lambda^a\,]   (\mbox{P}
e^{\int_0^{T_3}M(w)})_{\nu\sigma}^{ae}\ ,\label{gamma3}\\
\Gamma_4 
&=&\hskip-3pt {1\over8}\int_0^\infty\hskip-3pt dS\hskip-3pt
\int_0^S \hskip-3pt d\tau_\alpha\hskip-3pt 
\int_0^\infty \hskip-4pt dT_3 \oint[{\cal D}x]_S\hskip-3pt
\int_{\scriptstyle w(T_3)=x(\tau_\alpha) 
       \atop\scriptstyle w(0)=x(0)}[{\cal D}w]_{T_3}\,
{\dot x}_\nu(0){\dot w}_\rho(T_3)\nn \\
&\times& \mbox{Tr}[\, (\mbox{P}
e^{\int_0^{\tau_\alpha}M(x)})_{\mu[\sigma}\lambda^e
(\mbox{P}e^{\int_{\tau_\alpha}^S M(x)})_{\rho]\mu}
\lambda^a\,] (\mbox{P}
e^{\int_0^{T_3}M(w)})_{\nu\sigma}^{ae}\ ,\label{gamma3a}
\eeqa
where the indices $\rho$ and $\sigma$ in $\Gamma_3$ are 
anti-symmetrized by the lowercased symbol ${}_{[\cdots]}$ 
(Note that we used $x(0)=x(S)$). 
For the 1PR parts, we can proceed similarly with 
\beqa
&&(F \lambda^a G\lambda^b H)^{ab}=\mbox{Tr}[F\lambda^a]\,G^{ab}\,
\mbox{Tr}[\lambda^bH]\ ,\\
&&\mbox{Tr}[\,{\bar\Delta}_{\mu\nu}(y_1,y_2)\lambda^a\,]= 
-\mbox{Tr}[\,{\bar\Delta}_{\nu\mu}(y_2,y_1)\lambda^a\,]\ ,\qquad 
\mbox{Tr}[\,{\bar\Delta}_{\mu\mu}(y,y)\lambda^a\,]=0\ , 
\eeqa
and we obtain
\beqa
\Gamma_R &\equiv& \Gamma_1^R+\Gamma_2^R \nn\\
&=& {1\over8}\int_0^\infty dS \int_0^\infty dT 
\int_0^\infty dT_3 \oint[{\cal D}x]_S \oint[{\cal D}z]_T
\int_{\scriptstyle w(T_3)=z(T) 
       \atop\scriptstyle w(0)=x(S)}[{\cal D}w]_{T_3}\, \nn\\
&\times& ( {\dot w}_\mu(0){\dot w}_\sigma(T_3)  
          +{\dot x}_\mu(0){\dot z}_\sigma(T)) \nn\\
&\times& \mbox{Tr}[\,
(\mbox{P}e^{\int_0^S M(x)})_{\nu\mu} \lambda^a\,]
(\mbox{P}e^{\int_0^{T_3}M(w)})_{\nu\rho}^{ae}
\mbox{Tr}[\,(\mbox{P}e^{\int_0^T M(z)})_{\sigma\rho}\lambda^e\,]\ .
\label{gammar}
\eeqa
We shall verify the conjecture in this way: we show that 
$\Gamma^{(1)}_3$ and $\Gamma^{(2)}_3$ survive as singular integrands 
of $\Gamma_1$ and $\Gamma^R_1$ in the limit $T_3\ra0$, while 
$\Gamma_2$ and $\Gamma^R_2$ do not contribute. 

Let us first consider the 1PI case \eq{shrink1}. For the purpose of 
understanding the shrinking limit, we have only to analyze a free 
propagator; both edges of a shrinking propagator already involve 
four gluon lines, hence there is no need to take external gluon 
lines into account. The vacuum diagrams suffice since the short 
distance behavior in the vicinities of the two vertices is 
important. Actually the effect of external lines can be evaluated 
by insertions of vertex operators afterward. On these grounds it 
is sufficient to observe the $\phi^3$ world-line Green 
functions~\cite{SSphi,HTS,RS1}. The necessary two-loop world-line 
Green functions~\footnote{
One can equally well work with another representation of these 
Green functions~\cite{SSphi}. In that case, one should pay attention 
to the sign (direction) of each $\tau$ parameter~\cite{HTS}. 
The $\tau_\beta$ is revived here for convenience of presentation.} 
are as follows~\cite{HTS}:
\beqa
G^{(1)}_{ww}(\tau_1,\tau_2)&=&<w(\tau_1)w(\tau_2)>\, =\, 
|\tau_1-\tau_2|-{(\tau_1-\tau_2)^2\over 
T_3+G_B(\tau_\alpha,\tau_\beta)}\ ,  \label{g33}\\
G^{(1)}_{wx}(\tau_1,\tau_2)&=&<w(\tau_1)x(\tau_2)>\label{g30}\\
&=& G^{(1)}_{xx}(\tau_\beta,\tau_2) +
{1\over T_3+G_B(\tau_\alpha,\tau_\beta)}
(T_3\tau_1-\tau_1^2+\tau_1[\,G_B(\tau_2,\tau_\alpha)
-G_B(\tau_2,\tau_\beta)\,])\ ,\nn
\eeqa
where $G_B$ and $G^{(1)}_{xx}$ are the one-loop~\cite{St,poly} and 
two-loop~\cite{SSphi} world-line Green functions on the $x(\tau)$ 
loop (length $S$): 
\beqa
&&\hskip-50pt G_B(\tau_1,\tau_2)
       =|\tau_1-\tau_2|-{(\tau_1-\tau_2)^2\over S}\ ,\\
&&\hskip-50pt G^{(1)}_{xx}(\tau_1,\tau_2) 
= G_B(\tau_1,\tau_2)-{1\over4}
{(G_B(\tau_1,\tau_\alpha)-G_B(\tau_1,\tau_\beta)
-G_B(\tau_2,\tau_\alpha)+G_B(\tau_2,\tau_\beta))^2
\over T_3+G_B(\tau_\alpha,\tau_\beta) }\ .\label{Gxx}
\eeqa
We evaluate 
\beqa
&&\hskip-40pt <{\dot w}(0){\dot w}(T_3)> = 
\der_1\der_2 G^{(1)}_{ww}(\tau_1,\tau_2)\Bigr|_{\tau_1=0,\tau_2=T_3}
=2\delta(T_3)+{2\over T_3+G_B(\tau_\alpha,\tau_\beta)}\ , 
\label{wwdot} \\
&&\hskip-40pt <{\dot x}(0){\dot w}(T_3)> = 
\der_1\der_2 G^{(1)}_{wx}(\tau_1,\tau_2)\Bigr|_{\tau_1=T_3,\tau_2=0}
={-2|\tau_\beta-\tau_\alpha|\over 
            S(T_3+G_B(\tau_\alpha,\tau_\beta))}\ ,\label{wxdot}
\eeqa
where we have taken account of the absolute value of 
$\tau_\beta-\tau_\alpha$ in the second quantity \eq{wxdot}, 
since the Green function \eq{g30} is in fact defined for the 
ordering $\tau_\alpha<\tau_\beta$ ~\cite{HTS}. 
These yield the following relation:
\beq{pinch}
<({\dot w}_\mu(0)-{\dot x}_\mu(0)){\dot w}_\rho(T_3)> = 
2g_{\mu\rho}[\, \delta(T_3) +
{ S+|\tau_\alpha-\tau_\beta| \over
S(T_3+G_B(\tau_\alpha,\tau_\beta))}\,]\ .
\eeq
The first term on the r.h.s. in Eq.\eq{pinch} provides a singular 
term in the $T_3$ integrand, while the second term contributes to a 
regular quantity in the $T_3$ integrand, since we can not take both 
of $T_3\ra0$ and $\tau_\beta\ra\tau_\alpha$ simultaneously. 
(If one takes both of them, one of the loops shrinks to a point). 
The singularity $\delta(T_3)$ comes only from the 
quantity~\eq{wwdot}, and the shrinking conjecture~\eq{shrink1} 
clearly corresponds to this singularity in the limit $T_3\ra0$ 
(with $|\tau_\beta-\tau_\alpha|$ kept finite). Apparently 
$\Gamma_4$ does not contribute the singularity 
(because of Eq.\eq{wxdot}). 
 
The meaning of the other term in Eq.\eq{pinch} is the following. 
Let us take $\tau_\beta\ra\tau_\alpha$ in the r.h.s. of 
Eq.\eq{pinch}. It reduces to 
\beq{*}
\mbox{R.H.S. of}\quad \eq{pinch} \quad\ra\quad
2g_{\mu\rho}\left(\,\delta(T_3)+{1\over T_3}\,\right)\ .
\eeq
Note that this is the same form as the second derivative 
$\der_1\der_2G_B(\tau_1,\tau_2)$ of the one-loop Green function 
on a loop of length $T_3$. The $\delta$-function is the singular 
term relevant to the shrinking limit $T_3\ra0$ as mentioned above. 
The second term also seems to be singular, however it is not 
(Remember $T_3\not=0$). In the situation of 
$\tau_\beta\ra\tau_\alpha$, the $w(\tau)$ path integral becomes 
a loop integral (q.v. \eq{gamma3}) and this $T_3^{-1}$ factor 
simply describes a part of the usual loop integral measure: 
\beq{*}
\lim_{\epsilon\ra0}
\int_\epsilon^\infty {dT_3\over T_3} \oint [{\cal D}w]_{T_3}\ .
\eeq
Therefore we can embed the $\Gamma_3^{(1)}$ into $\Gamma_3$ 
as the integrable edge singularity at $T_3=0$ ($\epsilon=0$) 
produced by the Wick contraction of two vertices \eq{wwdot}. 

In the case of the 1PR function \eq{gammar}, we expect the 
similar shrinking \eq{shrink2} takes place. 
Since $x(\tau)$ and $z(\tau)$ can not approach each other, 
there is no singular term created from $\Gamma_2^R$. 
This non-singular situation itself is common in $\Gamma_2$ and 
$\Gamma_2^R$, and the position of $\tau_\alpha$ (whether 
$\tau_\alpha<\tau_\beta$ or $\tau_\beta<\tau_\alpha$) is not 
an important issue (see Fig. 3(b)). Hence the folding 
diagram piece Fig. 3(a) does not yield any singularity, and 
neither does the diagram Fig. 4(b). 
The only singular part comes from the ${\dot w}(0){\dot w}(T_3)$ 
contraction originated in $\Gamma_1^R$. To see this, the free (open) 
bosonic two-point function suffices: 
\beq{tree}
\der_1\der_2<w_\mu(\tau_1)w_\sigma(\tau_2)>^{tree}
\Bigr|_{\tau_1=0,\tau_2=T_3} = 
g_{\mu\sigma}\der_1\der_2|\tau_1-\tau_2|\Bigr|_{\tau_1=0,\tau_2=T_3} 
=2 g_{\mu\sigma}\delta(T_3) \ .
\eeq
Thus one can embed $\Gamma^{(2)}_3$ into the edge singularity 
of $\Gamma_R$, with obtaining the relation \eq{shrink2}. 
It is also easy to understand Eqs.\eq{rep2}, \eq{lim1}, \eq{lim2} 
due to Eqs.\eq{dotform} and \eq{tree}: 
\beq{*}
\lim_{T_3\ra0}D^{ai}_\mu(y_1)D^{ej}_\nu(y_2)K^{kl}_{\rho\sigma}
(y_1,y_2;0,T_3)=2({i\over2})^2\delta^{ai}\delta^{ej}\delta^{kl}
g_{\mu\nu}g_{\rho\sigma}\delta(y_1-y_2)\ .
\eeq
The factor ${1\over4}$ in Eq.\eq{lim2} is derived from the 
normalization ${1\over2}$ in the formula~\eq{dotform}, and the 
factor 2 in Eq.\eq{lim1} is the coefficient 2 in front of the 
$\delta(T_3)$ singularity. 

In this section, we have derived the shrinking relations 
\eq{shrink1} and \eq{shrink2} in paraphrase by world-line 
language: in other words we showed that $\Gamma^{(i)}_3$ can be 
embedded into $\Gamma_3$ and $\Gamma_R$ as the edge singularities 
of the $T_3$ integrals. We hence  have only three compact 
representations \eq{gamma3}, \eq{gamma3a} and \eq{gammar}. However, 
remember that a simpler alternative set of compact representations 
is of course to have $\Gamma_3$, $\Gamma_4$ and $\Gamma^{(2)}_3$ 
(see Eqs.\eq{W32} or \eq{W32a}).

\section{The pseudo-abelian case}\label{sec6}
\setcounter{section}{6}
\setcounter{equation}{0} 
\indent

We examine how the results obtained in Section~\ref{sec5} can be 
simplified in the $su(2)$ pseudo-abelian case with constant field 
strength. We then derive a general formula for the two-loop 
Euler-Heisenberg type action in this case. The setting is the 
following~\cite{RSS}: We assume the particular decomposition
\beqa
A^a_\mu = {\cal A}_\mu n^a\ ,\qquad 
F^a_{\mu\nu} = {\cal F}_{\mu\nu}n^a\ , \\
\qquad M_{ij}(x) = {\cal M}(x) (\lambda^a n^a)_{ij}\ ,
\eeqa
where all the color vectors are chosen to be proportional to a 
unit color vector; for instance, $\vec{n}=(n^1,n^2,n^3)$ for 
$su(2)$. The quantities ${\cal A}_\mu$, ${\cal F}_{\mu\nu}$, 
${\cal M}$ are all commuting quantities in color space, and we 
further assume the following relation for the commuting gauge field: 
\beq{*}
{\cal A}_\mu{\dot x}^\mu 
= {1\over2}x^\mu {\cal F}_{\mu\nu}{\dot x}^\nu\ .
\eeq

In the following we also use the brief notation
\beq{path}
{\cal P}_{\mu\nu}(i) = \left(
\mbox{P}\,\exp[\int M(x_i)d\tau]\right)_{\mu\nu}\ ; 
\qquad i=1,2,3 \ ,
\eeq
where $i$ stands for the $i$th path ordered exponential as they 
appear from left to right in Eqs.\eq{EA} and \eq{EAR}. Also the 
integration ranges should be understood in the same way. Let us 
consider the $su(2)$ case. Defining the following bases:  
\beq{*}
{\cal T}_- \equiv \lambda^a n^a\ , \qquad 
{\cal T}_+\equiv ({\cal T}_-)^2\ ,
\qquad {\cal I}=diag(1,1,1)-{\cal T}_+\ ,
\eeq
we can expand the path ordered exponential \eq{path} on these:
\beq{*}
{\cal P}_{\mu\nu}(i)={\cal P}^+_{\mu\nu}(i){\cal T}_+ 
+ {\cal P}^-_{\mu\nu}(i){\cal T}_- + \delta_{\mu\nu}{\cal I}\ ,
\eeq
where the coefficients are given by 
\beq{*}
{\cal P}^+(i) =\cosh(\int {\cal M}(x_i)d\tau)\ ,\qquad 
{\cal P}^-(i) =\sinh(\int {\cal M}(x_i)d\tau)\ .\qquad 
\eeq
We also define the following quantities just for compactness of 
presentation:
\beqa
{\cal P}^{\pm}(i,j)_{\mu\nu}
&\equiv& ({\cal P}^+(i){\cal P}^+(j))_{\mu\nu} 
\pm ({\cal P}^-(i){\cal P}^-(j))_{\mu\nu}  \nn\\
&=& \Bigl(\cosh(\int {\cal M}(x_i)d\tau \pm \int{\cal M}(x_j)d\tau)
\Bigr)_{\mu\nu}\ .
\eeqa

After some calculation by using the formulae 
\beq{*}
\mbox{Tr}[\,{\cal I}\lambda^e{\cal T}_\pm \lambda^a\,]
({\cal T}_\pm)^{ae}=2\ ,
\qquad 
\mbox{Tr}[\,{\cal T}_\pm \lambda^e{\cal T}_\pm \lambda^a\,]
({\cal I})^{ae}=2 \qquad\mbox{etc.}
\eeq
we obtain 
\beqa
\Gamma_3\dand=\dand-{1\over4}\int dS dT_3 
d\tau_\alpha 
\oint[{\cal D}x]_S \int_{\scriptstyle w(T_3)=x(\tau_\alpha) 
\atop\scriptstyle w(0)=x(0)} [{\cal D}w]_{T_3}\,
({\dot w}_\mu(0)-{\dot x}_\mu(0)){\dot w}_\nu(T_3)  \nn \\
\dand\times\dand\Bigl\{\, \Bigl({\cal P}^-(-1,2) 
+ {\cal P}^+(-1,3) + {\cal P}^+(2,3) \Bigr)_{\nu\mu}\hskip-2pt 
- {\cal P}_{\nu\mu}^+(2)\,
\mbox{Tr}_L[\,{\cal P}^+(1)+{\cal P}^+(3)\,] \nn\\
\dand-\dand {\cal P}_{\nu\mu}^-(2)\,
\mbox{Tr}_L[\,{\cal P}^-(1) + {\cal P}^-(3)\,]
-\delta_{\nu\mu}\mbox{Tr}_L {\cal P}^+(-1,3)\,
\Bigr\}\ ,\label{PG3}
\eeqa
where $\mbox{Tr}_L$ means the trace w.r.t. the Lorentz indices. 
The minus symbols of the first arguments in ${\cal P}^-(i,j)$ 
mean that the direction of their paths are reverted by the changes 
of $\tau$ directions. Also, applying the following formulae:
\beqa
&&\mbox{Tr}[\,{\cal P}_{\mu\nu}(1)\lambda^a 
{\cal P}_{\rho\sigma}(2)\lambda^a\,]=
2\Bigl(\, {\cal P}^+_{\mu\nu}(1)\delta_{\rho\sigma} 
+ \delta_{\mu\nu}{\cal P}^+_{\rho\sigma}(2) + 
\sum_{\kappa=\pm}{\cal P}^\kappa_{\mu\nu}(1)
{\cal P}^\kappa_{\rho\sigma}(2)  \,\Bigr)\ ,\\
&&\mbox{Tr}[\,{\cal P}_{\mu\nu}(1)\lambda^a\,]
\mbox{Tr}[\,{\cal P}_{\rho\sigma}(2)
\lambda^a\,]= 4{\cal P}^-_{\mu\nu}(1){\cal P}^-_{\rho\sigma}(2)\ ,\\
&&\mbox{Tr}[\,{\cal P}_{\mu\nu}(1)\lambda^a\,]
({\cal P}_{\nu\rho}(3))^{ae}
\mbox{Tr}[\,{\cal P}_{\rho\sigma}(2)\lambda^e\,] 
= 4{\cal P}^-_{\mu\nu}(1){\cal P}^-_{\nu\sigma}(2)\ ,
\eeqa
we obtain $\Gamma^{(i)}_3$ and the reducible function 
\eq{gammar} as follows: 
\beqa
\Gamma_3^{(1)}&=&
-{1\over2}\int dSd\tau_\alpha \oint
_{\scriptstyle x(0)=x(\tau_\alpha)} 
\hskip-30pt[{\cal D}x]_S\,\, \Bigl\{\,
-(\mbox{Tr}_L{\cal P}^+(1))(\mbox{Tr}_L{\cal P}^+(2))
-(\mbox{Tr}_L{\cal P}^-(1))(\mbox{Tr}_L{\cal P}^-(2)) \nn\\
&&\hskip60pt+\mbox{Tr}_L[\, (1-D)({\cal P}^+(1)+{\cal P}^+(2)) 
+{\cal P}^-(-1,2)\,]
\,\Bigr\} \ ,\label{PG31}\\
\Gamma_3^{(2)} &=&
\int dSdT\oint[{\cal D}x]_S[{\cal D}z]_T\, 
\mbox{Tr}_L[\,{\cal P}^-(1){\cal P}^-(2)\,]\ , \label{PG32}\\
\Gamma_R&=&
{1\over2}\int dSdTdT_3\oint[{\cal D}x]_S\oint[{\cal D}z]_T
\int_{\scriptstyle w(T_3)=z(T) \atop\scriptstyle w(0)=x(S)}
[{\cal D}w]_{T_3}\, \nn\\
&&\hskip60pt\times
({\dot w}_\mu(0){\dot w}_\nu(T_3)+{\dot x}_\mu(0){\dot z}_\nu(T_3))
\,({\cal P}^-(2){\cal P}^-(1))_{\nu\mu}\ .\label{PGR}
\eeqa
With the replacements 
\beq{*}
{\dot w}_\mu(0){\dot w}_\nu(T_3) \quad\ra\quad 
2g_{\mu\nu}\delta(T_3)\ ,\qquad\mbox{\rm{otherwise}}\ra 0\ ,
\eeq
it is easy to see that the previous shrinking limits hold: 
\beq{*}
\Gamma_R \quad\tlim{T_3}\quad \Gamma^{(2)}_3\ ,\qquad 
\Gamma_3 \quad\tlim{T_3}\quad \Gamma^{(1)}_3\ .
\eeq
Here is a remark. The r.h.s. of \eq{PG3} is not a symmetric 
expression in $1\leftrightarrow2$ exchange, just because of 
the fixing $\tau_\beta=0$. One may of course take an average 
to have the symmetric expression if any strong reason exists. 
The other quantities are symmetric as seen in 
Eqs.\eq{PG31}-\eq{PGR}.

Now, let us derive a general formula for an Euler-Heisenberg 
type action. We concentrate on the $\Gamma_3$ part, since we 
intend to discuss a general strategy only, and the 
$\Gamma^{(2)}_3$ part is straightforward from the one-loop 
action~\cite{RSS} (see Eq.\eq{PG32}). We assume 
${\cal F}_{\mu\nu}$ to be a constant matrix, and introduce the 
``symmetric'' type world-line representation with 
splitting $S=T_1+T_2$: 
\beq{*}
\int\hskip-3pt{dS\over S}dT_3d\tau_\alpha d\tau_\beta\oint
[{\cal D}x]_S\hskip-3pt \int_{\scriptstyle 
w(T_3)=x(\tau_\alpha) \atop\scriptstyle w(0)=x(\tau_\beta)}
[{\cal D}w]_{T_3} 
=\hskip-3pt\int\hskip-3pt d^Dy_1d^Dy_2\hskip-3pt\int\hskip-3pt 
dT_1dT_2dT_3 \prod_{a=1}^3 
\int_{\scriptstyle x_a(T_a)=y_2 \atop\scriptstyle x_a(0)=y_1}
[{\cal D}x_a]_{T_a}\ . 
\eeq
Every term in \eq{PG3} has one free internal line 
(without background field) out of three lines $a=1,2,3$. 
Denoting the free line label as $b$, we consider the following 
general term $\Gamma^{(b)}_{EH}$ 
for the linear combination \eq{PG3}: 
\beqa
\Gamma^{(b)}_{EH} \dand\define\dand \int \prod_{a=1}^3{}^{'} 
dT_a [{\cal D}x_a]_{T_a}
V_{\mu\nu}\Bigl(
 \exp[ \int_0^{T_a} {\cal M}(x_a)d\tau_a]\Bigr)_{\nu\mu} \nn\\
\dand=\dand\hskip-2pt \int \prod_{a=1}^3{}^{'}
\hskip-3pt dT_a {\cal D}x_a 
\Bigl(e^{2i{\cal F}T_a}\Bigr)_{\nu\mu} 
V_{\mu\nu} \exp[\,-{1\over4}\sum_{a=1}^3{}^{'}\hskip-3pt
\int_0^{T_a} ({\dot x}^2_a + 
2i \kappa^a x^{\mu}_a{\cal F}_{\mu\nu}
{\dot x}^{\nu}_a)d\tau_a\,] \ ,
\label{GbEH}
\eeqa
where $\kappa^a$ ($a\not=b$) is either of $\pm1$, 
\beq{*}
V_{\mu\nu}=\Bigl( {\dot x}_3^\mu(0)-{\dot x}_1^\mu(0)\Bigr)
x_3^\nu(T_3) \ ,
\eeq
and the primes on $\prod$ and $\sum$ denote to set 
${\cal F}_{\mu\nu}=0$ in the $x_b$ and $\tau_b$ integrals. 
In this paper, we omit the sign factor $\kappa^a$ for simplicity, 
since it is not difficult to revive it by rescaling ${\cal F} 
\ra \kappa{\cal F}$. After performing the path integrals 
(see Appendix C for details), the action $\Gamma_{EH}^{(b)}$ 
takes the following form: 
\beq{*}
\Gamma^{(b)}_{EH}=\Bigl(
\prod_{a=1}^3{}^{'}\int dT_a (e^{2i{\cal F}T_a})
\Bigr)_{\nu\mu}{\cal N}_b <V_{\mu\nu}>'\ ,
\eeq
where $<V_{\mu\nu}>'$ is an expectation value in the action 
\beq{*}
S^{(b)}=-{1\over4}\sum_{a=1}^3{}^{'}\int_0^{T_a}
({\dot x}_a^2 +2ix_a{\cal F}{\dot x}_a)d\tau_a\ ,
\eeq
and ${\cal N}_b$ is the path integral determinant factor
\beq{*}
{\cal N}_b= (4\pi)^{D\over2}
\Bigl(\prod_{a=1}^3(4\pi T_a)^{-{D\over2}}\Bigr)
\mbox{Det}_L^{-1/2}[\,\sum_a{}^{'}{\cal F}\cot({\cal F}T_a)\,]
\prod_a{}^{'}\mbox{Det}_L^{-1/2}
[\,{\sin({\cal F}T_a)\over{\cal F}T_a}\,]\ .
\eeq 
The quantity $<V_{\mu\nu}>'$ can be evaluated by the following 
world-line two-point correlator: 
\beq{Green}
{\cal G}_{\mu\nu}(\tau_a,\tau'_c)=
-\delta_{ac}{\tilde G}_{\mu\nu}(\tau_a,\tau'_c) +2\Bigl([\,
\sum{}^{'}{\cal F}\cot({\cal F}T_a)]^{-1}\Bigr)_{\rho\sigma}
\Bigl({e^{2i{\cal F}\tau_a}-1\over e^{2i{\cal F}T_a}-1}-{1\over2} 
\Bigr)_{\rho\mu}
\Bigl({e^{2i{\cal F}\tau'_c}-1\over e^{2i{\cal F}T_c}-1}-{1\over2} 
\Bigr)_{\sigma\nu}\ ,
\eeq
where 
\beq{Gcombi}
{\tilde G}_{\mu\nu}(\tau,\tau')= G^a_{\mu\nu}(\tau,\tau')
-G^a_{\mu\nu}(\tau,0)-G^a_{\mu\nu}(0,\tau')
\eeq
with
\beq{green}
G^a_{\mu\nu}(\tau,\tau')=
\left\{ \begin{array}{ll}
\delta_{\mu\nu}G^a_B(\tau,\tau')
=\delta_{\mu\nu}[\,|\tau-\tau'|-{(\tau-\tau')^2\over T_a}\,]
&\quad \mbox{(for $a$ on a free line)}\\ 
\Bigl[
{1\over2{\cal F}^2}\Bigl({{\cal F}\over\sin({\cal F}T_a)}
e^{-i{\cal F}T_a\der_\tau G^a_B(\tau,\tau')}
+i{\cal F}\der_\tau G^a_B(\tau,\tau')
-{1\over T_a}\Bigr)\Bigr]_{\mu\nu} &\quad \mbox{(otherwise)}\ .  
\end{array}\right.
\eeq

\section{The ghost loop}\label{sec7}
\setcounter{section}{7}
\setcounter{equation}{0} 
\indent

The two-loop contribution from the ghost part can be easily 
extracted from the generating functional \eq{ZA}:
\beq{Gamma4}
\Gamma_5=-i\int dy_1dy_2\mbox{Tr}\ln(D^2-D\beta f){i^3\over2}
\fder_\beta{\bar\Delta}\fder_\beta\Bigr|_{\alpha=0,\beta=0}\ .
\eeq
It is convenient to represent the $D\beta f$ term as 
\beq{*}
(D^2-D\beta f)^{ac}=(D^2)^{ac}-i\rcov{\mu}{ba}
\beta_\mu^d(\lambda^d)_{bc}\ ,
\eeq
and also to have the formal analogies of Eqs.\eq{path1} and 
\eq{path2} without representing the right derivative 
$\rcov{\mu}{\,\,}$ parts in terms of a world-line field:
\beqa
&&\hskip-40pt\mbox{Trln}(D^2-D\beta f)\,=\,-\int_0^\infty{dS\over S}
\oint{\cal D}x\exp[-\int_0^S{1\over4}{\dot x}^2(\tau)d\tau]
(\mbox{Pexp}\int_0^S{\bar N}[x(\tau)]d\tau)^{aa}\ ,
\label{path3}\\
&&\hskip-40pt{\bar\Xi}^{ab}(y_1,y_2)=\int_0^\infty d(\tau_2-\tau_1)
\int_{\scriptstyle x(\tau_2)=y_2 
        \atop\scriptstyle x(\tau_1)=y_1}{\cal D}x
e^{-\int_{\tau_1}^{\tau_2}{1\over4}{\dot x}^2(\tau)d\tau}
(\mbox{Pexp}\int_{\tau_1}^{\tau_2}{\bar N}[x(\tau)]d\tau)^{ab}\ ,
\label{path4}
\eeqa
where
\beq{*}
{\bar N}_{ab}[x(\tau)]=-iA^c_\mu{\dot x}^\mu\,(\lambda^c)_{ab}
-i\rcov{\mu}{ca}\beta_\mu^d(\lambda^d)_{cb}\ .
\eeq
Here we have a definite reason of not expressing the right 
derivative $\rcov{\mu}{\,\,}$ in terms of the world-line field at 
this stage. This kind of derivative is understood to be 
artificially exponentiated as a consequence of the introduction 
of auxiliary field $\beta$, since the $\beta$-derivatives in 
Eq.\eq{Gamma4} get the right derivatives $\rcov{\,\,}{\,\,}$ down 
from the exponent. Thus the $\rcov{\,\,}{\,\,}$ acts on a boundary 
of the path integral, and it will be replaced by a world-line field 
according to formula \eq{dotform} after all. 

Substituting the expressions \eq{path2}, \eq{path3} and \eq{path4} 
in Eq.\eq{Gamma4}, we can perform the $\fder_\beta$ differentiations 
in the same way as done in Section~\ref{sec3b} with 
\beq{*}
{\fder {\bar N}^{ic}[x(\tau)]\over\fder\beta^a_\mu(y)}
=-i\rcov{\mu}{bi}(y)(\lambda^a)_{bc}\delta(y-x(\tau))\ .
\eeq
The only difference is that we have to cut open the loop integral 
to manifest the boundaries where the $\rcov{\,\,}{\,\,}$'s operate:
\beqa
&&\int_0^\infty{dS\over S}\int_0^S d\tau_\alpha \int_0^S d\tau_\beta 
\oint[{\cal D}x]_S\rcov{\nu}{\,\,}(y_2)\rcov{\mu}{\,\,}(y_1)
\delta(y_1-x(\tau_\beta))\delta(y_2-x(\tau_\alpha))\nn\\
&&\hskip20pt=\int d^Dy'_1d^Dy'_2\int_0^\infty dT_1\int_0^\infty dT_2 
\int_{\scriptstyle x(T_1)=y'_2 
        \atop\scriptstyle x(0)=y_1}[{\cal D}x]_{T_1}
\int_{\scriptstyle {\bar x}(T_2)=y'_1 
    \atop\scriptstyle {\bar x}(0)=y_2}[{\cal D}{\bar x}]_{T_2}\nn\\
&&\hskip20pt\times \rcov{\nu}{\,\,}(y'_2)\rcov{\mu}{\,\,}(y'_1)
\delta(y_1-{\bar x}(T_2))\delta(y_2-x(T_1))\Bigr|_{y'_i=y_1}\ .
\eeqa
The result is thereby 
\beqa
\Gamma_5&=&-{1\over2}\int_0^\infty dT_1 dT_2 dT_3
\int_{\scriptstyle x(T_1)=y'_2 
        \atop\scriptstyle x(0)=y_1}[{\cal D}x]_{T_1}
\int_{\scriptstyle {\bar x}(T_2)=y'_1 
        \atop\scriptstyle {\bar x}(0)=y_2}[{\cal D}{\bar x}]_{T_2}
\int_{\scriptstyle w(T_3)=x(T_1)
        \atop\scriptstyle w(0)={\bar x}(T_2)}[{\cal D}w]_{T_3}
\label{g4}\\
&\times&
(\mbox{P}e^{\int_0^{T_3}M(w)})^{ae}_{\mu\nu}
(\mbox{P}e^{\int_0^{T_1}N(x)})^{cj}
\rcov{\nu}{fj}(y'_2)(\lambda^e)_{fg}
(\mbox{P}e^{\int_0^{T_2}N({\bar x})})^{gi}
\rcov{\mu}{bi}(y'_1)(\lambda^a)_{bc}\Bigr|_{y'_i=y_i} \ ,\nn
\eeqa
where the $y'_i$ integrations are implicit for simplicity, and 
\beq{*}
N(x)={\bar N}(x)\Bigr|_{\beta=0}=-iA^a_\mu x_\mu \lambda^a\ .
\eeq
Then applying the formula \eq{dotform}, we finally reach 
the expression 
\beqa
\Gamma_5&=&{1\over8}\int_0^\infty dT_1 dT_2 dT_3
\int_{\scriptstyle x(T_1)={\bar x}(0) 
        \atop\scriptstyle x(0)={\bar x}(T_2)}[{\cal D}x]_{T_1}
[{\cal D}{\bar x}]_{T_2}
\int_{\scriptstyle w(T_3)=x(T_1)
        \atop\scriptstyle w(0)={\bar x}(T_2)}[{\cal D}w]_{T_3}\nn\\
&\times&
{\dot x}_\nu(T_1){\dot {\bar x}}_\mu(T_2)
(\mbox{P}e^{\int_0^{T_3}M(w)})^{ae}_{\mu\nu}
\,\mbox{Tr}\,[\,
(\mbox{P}e^{\int_0^{T_1}N(x)})\, \lambda^e\,
(\mbox{P}e^{\int_0^{T_2}N({\bar x})}) \lambda^a\,] \ .
\eeqa

In closing this section, several remarks are in order. (i) 
One may further replace 
${\dot x}_\nu(T_1){\dot {\bar x}}_\mu(T_2)$ with 
${\dot x}(\tau_\alpha){\dot x}(\tau_\beta)$ to have 
a closed path integral for $x$ like $\Gamma_1$. 
However, before doing this, one should keep in mind that 
no (pinching) singularity is caused by $\tau_\alpha\ra
\tau_\beta$, as expected from the ordinary Feynman rule. 
(ii) In the same way as the argument of Section~\ref{sec5}, 
this fact can easily be justified either from 
the symmetric world-line Green function~\cite{SSphi} 
or from the following loop type Green function~\cite{HTS}: 
\beq{*}
G^{(1)}_{x{\bar x}}(\tau,{\bar\tau})
=G_B(\tau,{\bar\tau})_{\tau<{\bar\tau}}-
{1\over T_3+G_B(\tau_\alpha,\tau_\beta)}
(\tau_\beta-{\bar\tau}-|\tau_\alpha-\tau_\beta|
{\tau-{\bar\tau}\over S})^2\ ,
\eeq
where ${\bar\tau}$ is the world-line parameter for the ${\bar x}$ 
field, and the ordering $\tau_\alpha<\tau_\beta$ is assumed 
by definition. Note that the first term $G_B(\tau,{\bar\tau})$ 
does not generate the singularity because of the ordering 
constraint $\tau<{\bar\tau}$, and 
\beq{*}
<{\dot x}(T_1){\dot{\bar x}}(T_2)>=
\der_\tau\der_{\bar\tau}\, G^{(1)}_{x{\bar x}}(\tau,{\bar\tau})
\Bigr|_{\tau=T_1,{\bar\tau}=T_2}
={2T_3\over S(T_3+G_B(\tau_\alpha,\tau_\beta)}\ .
\eeq
(iii) On the contrary, in the case of scalar loop, the Green 
function \eq{Gxx} generates the singularity, which is a 
realization of embedding contact interaction. 

\section{The fermion loop}\label{sec8}
\setcounter{section}{8}
\setcounter{equation}{0} 
\indent

In this section, we make a short remark on the fermion loop case. 
The fermion two-loop part can be extracted from 
the action \eq{action} as
\beq{Gamma5}
\Gamma_6=-i\int dy_1dy_2\mbox{Tr}\ln(\gamma^\mu(i{\hat D}_\mu
-\epsilon^a{\hat\lambda}^a)) {i^3\over2}
\fder_\epsilon{\bar\Delta}\fder_\epsilon 
\Bigr|_{\alpha=0,\epsilon=0}\ .
\eeq
Regarding the auxiliary field $\epsilon^a_\mu$ in the 
determinant in Eq.\eq{action} as a counterpart of classical 
background field in the covariant derivative, we have only to 
perform the shift $A^a_\mu \ra A^a_\mu-\epsilon^a_\mu$ in 
the usual world-line fermion loop formula, thus 
\beqa
&&\hskip-40pt\mbox{Trln}
(\gamma^\mu(i{\hat D}-\epsilon^a{\hat\lambda}^a)\,=\,-2\int_0^\infty
{dS\over S}
\oint{\cal D}x{\cal D}\psi\mbox{Tr}\mbox{P}\exp[
-\int_0^S(L_0 + {\bar M}_F)d\tau]\ , \label{floop}\\
&&\hskip-40pt{\bar\Delta}_F^{ij}(y_1,y_2)\define 
\int_0^\infty d(\tau_2-\tau_1)
\int_{\scriptstyle x(\tau_2)=y_2 
        \atop\scriptstyle x(\tau_1)=y_1}{\cal D}x{\cal D}\psi
(\mbox{Pexp}-\int_{\tau_1}^{\tau_2}(L_0+{\bar M}_F)d\tau)^{ij}\ ,
\label{sfprop}
\eeqa
where 
\beqa
L_0&=& {1\over4}{\dot x}_\mu^2+{1\over2}\psi^\mu{\dot\psi}_\mu 
+iA_\mu{\dot x}_\mu-i\psi^\mu F_{\mu\nu}\psi^\nu \ , \\
{\bar M}_F &=&  
-i\epsilon_\mu{\dot x}_\mu 
-\psi^\mu[\epsilon_\mu,\epsilon_\nu]\psi^\nu
+2i\psi^\mu(D^{ca}_\mu\epsilon^a_\nu)\psi^\nu {\hat \lambda}^c\ ,
\eeqa
and $\epsilon_\mu=\epsilon^a_\mu{\hat\lambda}^a$ etc. 
If we apply the derivative $\fder/\fder\epsilon^a_\mu$ 
to the loop \eq{floop}, we get a vertex
\beq{*}
{\fder {\bar M}_F(\tau) \over\fder\epsilon^a_\mu(y_1)}=
\delta(y_1-x(\tau)) \Bigl(\,
i({\dot x}_\mu - 2\psi^\nu D^{ca}_\nu \psi^\mu){\hat\lambda}^c 
+{\fder\over\fder\epsilon^a_\mu} 
\psi\cdot[\epsilon,\epsilon]\cdot\psi\,\Bigr) \ .\label{MFe}
\eeq
The first term on the r.h.s. of \eq{MFe} is pointed out in the 
abelian context in~\cite{SSphi}. The second term (the commutator 
term in \eq{MFe}) further contributes in a second derivative, 
and it becomes  
\beq{fpinch}
{\fder^2 {\bar M}_F(\tau) \over\fder\epsilon^b_\nu(y_2)
\epsilon^a_\mu(y_1)}=
\delta(y_1-x(\tau))\delta(y_2-x(\tau))
2\psi^\mu(\tau)[{\hat\lambda}^a,{\hat\lambda}^b]\psi^\nu(\tau)\ .
\eeq
It is worthwhile noting that this is equivalent to the pinching 
terms prescribed at the one-loop level by Strassler~\cite{St} 
(see also ~\cite{HTS2}): 
\beq{*}
Eq.\eq{fpinch} \sim {\cal O}_{ji}+{\cal O}_{ij}\ ,
\eeq
where $\sim$ means to remove the plane wave modes 
$\epsilon_n\exp[ik_nx(\tau_n)]$; $n=i,j$ in order to form a 
second loop by joining the two external lines, and 
\beq{*}
{\cal O}_{ji}=(-ig)^2 {\hat\lambda}^{a_j}
{\hat\lambda}^{a_i}\int_0^S
d\tau_j d\tau_i \delta(\tau_j-\tau_i)
2\epsilon_j\cdot\psi(\tau_j) \epsilon_i\cdot\psi(\tau_i)
e^{i(k_i+k_j)\cdot x(\tau_i)}\ .
\eeq

As in the case of QED~\cite{SSqed}, the two-loop effective 
action can be formulated very elegantly using world-line 
supersymmetry on the fermion loop: One has to substitute 
supervariables $X_\mu({\hat\tau})=x_\mu(\tau)+\sqrt{2}\theta
\psi_\mu(\tau)$ with ${\hat\tau}=(\tau,\theta)$ everywhere 
for $x_\mu(\tau)$ on the loop in the corresponding scalar case; 
in particular the superaction only contains the interaction 
term $DX_\mu({\hat\tau})A_\mu(X)$ 
($D=\der_\theta-\theta\der_\tau$), and the inserted gauge field 
propagator has a simple form in supervariables at its endpoints. 
The non abelian commutator pieces are then obtained 
automatically by a supersymmetric generalization of the 
ordering $\theta$-function~\cite{AT}. 

Given our formalism for pure Yang-Mills theory, it is 
also an interesting question if a world-line supersymmetric 
formulation of the interaction part of the action together 
with a (necessarily) supersymmetry breaking kinetic part 
can be used for a compact formulation.

\section{Conclusions}\label{sec9}
\setcounter{section}{9}
\setcounter{equation}{0} 
\indent

In this paper, we presented a method how to construct the 
two-loop effective action in terms of the bosonic world-line path 
integral representation in pure Yang-Mills theory, and we 
discussed the way how one of the ``eight figure'' sewing diagrams 
can be unified in the $\Phi^3$ type sewing diagram. The effective 
action is then summarized as 
\beq{final}
\Gamma^{2-loop} = \Gamma_3 + \Gamma_4 + \Gamma^{(2)}_3 \ , 
\eeq
and the additional term $\Gamma^{(2)}_3$ is related to one of the 
1PR sewing contributions (vid. Fig. 4(a)). 

In Section~\ref{sec2}, we explained the method how to obtain the 
two-loop analogue of the trace-log formula in the case of (full) 
Yang-Mills theory, using the background field and auxiliary field 
methods. The pure Yang-Mills part of the generating functional 
\eq{pureac} resembles the $\phi^3$ theory case; i.e., the 
propagator ${\bar\Delta}$ consists of the free propagator, 
background field and auxiliary field terms. The whole generating 
functional \eq{ZA} contains the full information of how to join the 
trace-log loops and the gluon propagators at any loop order. 
The sewing method does not require the computation of a number of 
Wick contractions w.r.t. space-time fields, and organizes all such 
terms into a compact expression automatically (as seen in Appendix A). 
The world-line calculations are thus expected to be easier 
than those in the space-time Wick contraction method. 

In Section~\ref{sec3}, we performed this sewing procedure in the 
language of the world-line representation. We realized that the 
bosonic field representations \eq{path1} and \eq{path2} fit well 
with the sewing method. All the $\Gamma_i$ including 1PR parts 
coincide with the $\Gamma$'s (in Eq.\eq{agammas}) evaluated 
without world-line (path integral) representation. It is worth 
noting that the emerging ways of the triple products of 
propagators are totally different from each other. One comes from 
the cutting rule \eq{cut} of the path integral, and the other is 
from power series expansions. This is certainly a non-trivial 
observation showing how the purely bosonic world-line 
representations are integrated in the entire framework 
(with the help of the sewing method). 

In Section~\ref{sec5}, we derived the compact world-line formulae 
\eq{gamma3}, \eq{gamma3a} and \eq{W32} for the two-loop pure 
Yang-Mills theory. We showed that Eq.\eq{gamma3} includes the 
$\Gamma^{(1)}_3$ part as the edge singularity of the integrand 
of $\Gamma_3$, and as a result, the full effective 
action~\eq{final} is simply the sum of \eq{gamma3}, \eq{gamma3a} 
and \eq{W32}. We also found the similar relation between 
$\Gamma^{(2)}_3$ and $\Gamma_R$. The latter relation 
itself seems not to be useful, since having the crude expression 
of $\Gamma^{(2)}_3$ (q.v. Eq.\eq{W32}) is much simpler than the 
embedding into $\Gamma_R$. However it was certainly helpful to 
check the validity of the shrinking limit. In the $su(2)$ 
pseudo-abelian case with much simpler expressions, we observed 
these results more clearly in Section~\ref{sec6}. It is also 
interesting to note that the addition of the 1PR related 
part $\Gamma^{(2)}_3$ resembles the one-loop observation that 
we have to add and shrink 1PR tree vertices in the Bern-Kosower 
rules~\cite{BK}. 

We also wrote down the ghost part for the two-loop effective 
action in Section~\ref{sec7}. An interesting question is how to 
realize gauge independence~\cite{inde} by gathering the 
ghost loop $\Gamma_5$ and the pure Yang-Mills part \eq{final}. 
We still have a problem to obtain a path integral representation 
of the gluon propagator in an arbitrary gauge, however this 
could be avoided formally separating the propagator into 
Feynman gauge part plus others~\cite{prop}. This strategy works 
in the one-loop case, however the present status between 
$\Gamma_5$ and $\Gamma_3+\Gamma_4$ (plus $\Gamma^{(2)}_3$) 
seems still far away from the goal.  

Nevertheless, we now have the effective action, and we will be 
able to compute amplitudes with Fourier expanding the background 
fields (i.e. substituting plane wave modes as explained in 
Appendix B). This substitution determines a combinatorial 
factor, and the whole procedure is straightforward. After that, 
we will be able to compare our results with conventional 
calculations. Studying a connection to string theory would also 
be of interest \cite{vacuum,Kaj}. A non-abelian version of the 
Euler-Heisenberg Lagrangian in two-loop order~\cite{RSS,KS} 
is a possible further interest as well. In this case we should 
proceed to the linear combination of $\Gamma_{EH}^{(b)}$ to get 
a more detailed expression. We hope to address 
these issues in a future publication. 

\noindent
{\large\bf Acknowledgment}

We are grateful to M. Reuter for useful discussions and 
contribution to Section~\ref{sec2}. We also thank C. Zahlten 
for numerous comments and discussions. 

\appendix

\section*{Appendix A. Equivalence to Feynman diagram technique}
\label{ap1}
\setcounter{section}{1}
\setcounter{equation}{0}
\indent

We present an equivalence of the world-line path-integral formulae 
to the results with the conventional technique. First of all, 
without introducing the path-integral representations \eq{path1} 
and \eq{path2}, one can of course perform the functional 
$\fder_\alpha$ differentiations directly: with using the cyclicity 
of color trace, the anti-symmetry of $f^{abc}=i(\lambda^a)_{bc}$, 
and the following relation
\beq{*}
\Delta^{ab}_{\mu\nu}(1,2)\equiv \Delta^{ab}_{\mu\nu}(x_1,x_2)=
\Delta^{ba}_{\nu\mu}(x_2,x_1)\ .
\eeq
The results are organized as follows:
\beqa
&&\Gamma_1={1\over4}\gmnrs\int dx_1 dx_2
(D^{ai}_{\mu'}(x_1)D^{ej}_{\rho'}(x_2)
\Delta^{ij}_{\nu'\sigma'}(1,2))\mbox{Tr}[\,\Delta_{\nu\rho}(1,2)
\lambda^e\Delta_{\sigma\mu}(2,1)\lambda^a\,]\ ,\nn\\
&&\Gamma_1^R={1\over8}\gmnrs \int dx_1 dx_2 
(D^{ai}_{\mu'}(x_1)D^{ej}_{\rho'}(x_2)
\Delta^{ij}_{\nu'\sigma'}(1,1))\mbox{Tr}[\,\Delta_{\mu\nu}(2,2)
\lambda^a\,] \mbox{Tr}[\,\Delta_{\rho\sigma}(1,2)\lambda^e\,]\ ,\nn\\
&&\Gamma_2={1\over2}\gmnrs \int dx_1 dx_2 
D^{ai}_{\mu'}(x'_1)D^{ej}_{\rho'}(x'_2)
[\,\Delta_{\nu'\rho}(1',2)\lambda^e\Delta_{\sigma\mu}(2,1)\lambda^a
\Delta_{\nu\sigma'}(1,2')\,]^{ij} \ ,\nn \\
&&\Gamma_2^R={1\over2}\gmnrs \int dx_1 dx_2 
D^{ai}_{\mu'}(x'_1)D^{ej}_{\rho'}(x'_2)
[\,\Delta_{\nu'\mu}(1',1)\lambda^a\Delta_{\nu\rho}(1,2)\lambda^e
\Delta_{\sigma\sigma'}(2,2')\,]^{ij}\ ,\nn\\
&&\Gamma_3^{(1)}=-{1\over8}\gmnrs \int dx_1 dx_2 
g_{\mu'\rho'}g_{\nu'\sigma'}
\mbox{Tr}[\,\Delta_{\nu\rho}(1,2)\lambda^a\Delta_{\sigma\mu}(2,1)
\lambda^a\,]\ ,\nn\\
&&\Gamma_3^{(2)}=-{1\over16}\gmnrs \int dx_1 dx_2
g_{\mu'\rho'}g_{\nu'\sigma'}
\mbox{Tr}[\,\Delta_{\mu\nu}(1,1) \lambda^a\,] 
\mbox{Tr}[\,\Delta_{\rho\sigma}(2,2)\lambda^a\,]\ ,\label{agammas}
\eeqa
where $x'_i$; $i=1,2$, are introduced to express derivative's 
positions where to operate, and they should be set to be $x_i$ 
after the differentiations. Here a few remarks are in order. 
(i) One can see the consistency of the world-line representation 
\eq{EA} with these results if identifying the quantity \eq{path2} 
(with $\alpha^a_{\mu\nu}=0$) to be $\Delta^{ab}_{\mu\nu}$. 
(ii) The above results are not unique expressions. 
{}~For example, noticing the formulae 
\beqa
&&\mbox{Tr}[\,\Delta_{\mu\nu}(1,2)\lambda^a\,]= 
-\mbox{Tr}[\,\Delta_{\nu\mu}(2,1)\lambda^a\,]\ ,\\
&&(\lambda^a)_{ij}(\lambda^a)_{kl}=(\lambda^a)_{ik}(\lambda^a)_{jl}
-(\lambda^a)_{jk}(\lambda^a)_{il}\ ,\\
&&\gmnrs g_{\mu'\rho'}g_{\nu'\sigma'} 
= 2\delta^{\mu\nu}_{\rho\sigma}\ ,
\eeqa
we can also have the following expressions:
\beqa
\Gamma_3^{(2)}&=& -{1\over4}
\mbox{Tr}[\,\Delta_{\mu\nu}(1,1) \lambda^a\,] 
\mbox{Tr}[\,\Delta_{\mu\nu}(2,2)\lambda^a\,]\ , \label{W32a}\\
  &=& {1\over8}\gmnrs g_{\mu'\rho'}g_{\nu'\sigma'}
\mbox{Tr}[\,\Delta_{\sigma\rho}(1,1)\lambda^a
\Delta_{\mu\nu}(2,2)\lambda^a\,]\ .
\eeqa

Let us derive the 1PI parts of the above $\Gamma_i$ in another 
method; i.e. by Wick contracting quantum fields $Q^a_\mu$. 
We have the pure Yang-Mills action 
\beq{*}
{\cal L}_{YM}={\cal L}_0+{\cal L}_3+{\cal L}_4\ ,
\eeq
where ${\cal L}_0$ stands for the one-loop (kinetic) term 
of quantum gauge field, and the other terms are 
\beqa
&&{\cal L}_3=-f^{abc}Q^a_{\mu\nu}Q^b_\mu Q^c_\nu \ ;
\qquad Q^a_{\mu\nu}= (D_\mu Q_\nu)^a\ ,\\
&&{\cal L}_4=-{1\over4}(f^{abc}Q^a_\mu Q^c_\nu)^2\ .
\eeqa
The two-loop effective action can thus be obtained as 
\beqa
i\Gamma^{2-loop}&=&<i{\cal L}_4>_0 +
{1\over2}<i{\cal L}_3(x_1)i{\cal L}_3(x_2)>_0 \nn \\
&\equiv& iS_4 + iS_3\ ,
\eeqa
where $<X>_0$ means a correlation function evaluated in the one-loop 
action ${\cal L}_0$ (The space-time integrations are implicit). 
{}~First let us consider the $S_3\sim<{\cal L}_3{\cal L}_3>_0$ part. 
After Wick contracting by using the gluon propagator
\beq{QQ}
<Q^a_\mu(x_i)Q^b_\nu(x_j)>_{ij} =-i\Delta^{ab}_{\mu\nu}(i,j)\ ,
\eeq  
we split $S_3$ into the following two quantities 
specified by single and double contractions w.r.t. $Q^a_{\mu\nu}$:
\beq{*}
S_3={i\over2}<{\cal L}_3(x_1){\cal L}_3(x_2)>_0
=W_1+W_2\ ,
\eeq
where
\beqa
&&W_1=-{i\over2}f^{abc}f^{efg}\Bigl[\,
<Q^a_{\mu\nu}Q^e_{\rho\sigma}>_{12} - 
(\rho\leftrightarrow\sigma)\,\Bigr]
\Delta^{bf}_{\mu\rho}(1,2) \Delta^{cg}_{\nu\sigma}(1,2)\ ,\\
&&W_2={1\over2}f^{abc}f^{efg} <Q^a_{\mu\nu}Q^f_\rho>_{12} 
\Bigl(\,\Delta^{cg}_{\nu\sigma}(1,2) [\,<Q^b_\mu Q^e_{\rho\sigma}>_{12}
                   - (\rho\leftrightarrow\sigma)\,] \\
&&\hskip50pt+ \Delta^{bg}_{\mu\sigma}(1,2) [\,
<Q^c_\nu Q^e_{\rho\sigma}>_{12}
               - (\rho\leftrightarrow\sigma)\,]\Bigr)\ .\nn
\eeqa
After some rearrangement to have the antisymmetric symbols 
$\gmnrs$, we verify the equalities
\beq{*}
W_1=\Gamma_1\ ,\qquad\quad W_2=\Gamma_2\ .
\eeq

Next, let us consider the ${\cal L}_4$ part in the same way. 
We first divide $S_4$ into the following quantities 
(self and mutual contractions w.r.t. space-time coordinates):
\beqa
S_4 &=& -{1\over4}f^{abc}f^{efg}
<Q^b_\mu Q^c_\nu(x_1)Q^f_\rho Q^g_\sigma(x_2)>_0 
\delta^{ae}g^{\mu\rho}g^{\nu\sigma}\delta(x_1-x_2)\nn\\
&=& W_3^{(1)}+W_3^{(2)}\ ,
\eeqa
where the self-contraction part is given by
\beq{*}
W_3^{(2)}=-{1\over4}f^{abc}f^{efg}
\delta^{ae}g^{\mu\rho}g^{\nu\sigma}\delta(x_1-x_2)
<Q^b_\mu Q^c_\nu>_{11}<Q^f_\rho Q^g_\sigma>_{22}\ ,
\eeq
and the mutual-contraction part is given by 
\beqa
W_3^{(1)}&& =-{1\over4}f^{abc}f^{efg}
\delta^{ae}g^{\mu\rho}g^{\nu\sigma}\delta(x_1-x_2)\nn\\
&&\times\Bigl(\,
<Q^b_\mu Q^f_\rho>_{12} <Q^c_\nu Q^g_\sigma>_{12}
+<Q^b_\mu Q^g_\sigma>_{12}<Q^c_\nu Q^f_\rho>_{12}
\,\Bigr) \ .
\eeqa
We thus verify that 
\beq{*}
W_3^{(1)}=\Gamma_3^{(1)}\ ,\qquad\quad  W_3^{(2)}=\Gamma_3^{(2)}\ .
\eeq

The ghost part is evaluated as follows. The action is 
\beq{*}
{\cal L}_{FP}={\bar c}D^2c + f^{abc}c^a(D_\mu{\bar c})^bQ^c_\mu\ ,
\eeq
and the two-loop contribution is given by 
\beq{*}
W_5 = {i\over2}f^{abc}f^{efg}
<c^a(D_\mu{\bar c})^bQ^c_\mu(x_1) 
c^e(D_\nu{\bar c})^fQ^g_\nu(x_2)>\ ,
\eeq
which can be evaluated with the propagators \eq{QQ} and 
\beq{*}
<{\bar c}^ac^b>_{12}=-i\Xi^{ab}(1,2)\ .
\eeq
We thus obtain
\beq{aW4}
W_5 = {1\over2}\Delta^{ae}_{\mu\nu}(1,2)
(D_\mu(x_1)\Xi(1,2))^{bg}(\lambda^e)_{gf}
D^{fj}_\mu(x_2)\Xi^{cj}(1,2)(\lambda^a)_{cb}\ ,
\eeq
and this coincides with Eq.\eq{g4} identifying 
${\bar\Xi}^{ab}$ (with $\beta^a_\mu=0$) to be $\Xi^{ab}$:
\beq{*}
W_5 = \Gamma_5\ .
\eeq

\section*{Appendix B. General structure of gluon $N$-point functions}
\label{ap2}
\setcounter{section}{2}
\setcounter{equation}{0}
\indent

In this appendix, we explain a general prescription how to calculate 
the $N$-point proper functions in the (two-loop) pure Yang-Mills 
case. Let us start with the following function:
\beq{*}
\Gamma^{(0)}\define
\int d^Dy_1 d^Dy_2\prod_{a=1}^3\int dT_a
\int_{\scriptstyle x_a(T_a)=y_2 
        \atop\scriptstyle x_a(0)=y_1}[{\cal D}x_a]_{T_a}
\mbox{Pexp}[\,
\int_0^{T_a}M(x_a)d\tau^{(a)}\,]^{j_al_a}_{\mu_a\nu_a} 
\lambda(a)
\eeq
or equivalently 
\beq{type2}
\Gamma^{(0)}= \int_0^\infty dS dT_3 \int_0^S d\tau_\alpha
\oint[{\cal  D}x]_S
\int_{\scriptstyle w(T_3)=x(\tau_\alpha)
        \atop\scriptstyle w(0)=x(S)}[{\cal D}w]_{T_3}
{\cal P}_{\mu_1\nu_1}(x)
\lambda^{a_\alpha}
{\cal P}_{\mu_2\nu_2}(x)
\lambda^{a_\beta}
{\cal P}_{\mu_3\nu_3}(w)
\eeq
where we have defined $\lambda(1)=\lambda^{a_\alpha}$,  
$\lambda(2)=\lambda^{a_\beta}$, and $\lambda(3)=1$, and the 
notation ${\cal P}_{\mu\nu}$ is defined in Eq.\eq{path}.   

To calculate the corresponding $N$-point proper Green functions 
($\equiv\Gamma^{(0)}_N$), we expand the background 
(interaction) terms 
\beq{*}
\Bigl( \mbox{P}e^{\int_0^{T_a} M d\tau^{(a)}}\Bigr)^{j_al_a}
_{\mu_a\nu_a}
=\sum_{N_a=0}^\infty (P_{N_a})^{j_al_a}_{\mu_a\nu_a}
\eeq
where
\beq{*}
P_{N_a}= \int_0^{T_a} d\tau^{(a)}_{N_a}\int_0^{\tau^{(a)}_{N_a}}
d\tau^{(a)}_{N_a-1}\cdots\int_0^{\tau^{(a)}_2}d\tau^{(a)}_1
\, M(\tau^{(a)}_1)\cdots M(\tau^{(a)}_{N_a})\ ,
\eeq
and insert the plane wave modes \eq{FT} in this expression. 
Then $P_{N_a}$ becomes multiple (ordered) integrals of 
the following quantity: 
\beq{*}
{\tilde V}_j(\tau) = -ig\lambda^{a_j}
\Bigl[\,\epsilon^j \cdot {\dot x}\delta_{\mu\nu}
-2i(k^j_\mu \epsilon^j_\nu 
-k^j_\nu \epsilon^j_\mu)\,\Bigr]\,e^{ik^j\cdot x(\tau)} 
+{\cal O}(g^2)\ .
\eeq
Here we neglect the higher order term, since we expect that 
it will be evaluated later with the pinching techniques 
(vid. Refs.\cite{St,HTS2,AT}). 

As usual~\cite{St,combi}, we are allowed to perform the following 
replacement by virtue of the total momentum conservation law 
(concerning the external legs):  
\beq{*}
(P_{N_a})^{jl}_{\mu\nu} \ra
\sum_{\sigma(N_a)}
\int_0^{T_a} d\tau^{(a)}_{i_{N_a}}
\int_0^{\tau^{(a)}_{i_{N_a}}} d\tau^{(a)}_{i_{N_a-1}}  \cdots 
\int_0^{\tau^{(a)}_{i_2}}  d\tau^{(a)}_{i_1}
\,\Bigl(\, {\tilde V}_{i_1}(\tau^{(a)}_{i_1})\cdots 
{\tilde V}_{i_{N_a}}(\tau^{(a)}_{i_{N_a}})\Bigr)^{jl}_{\mu\nu}\ ,
\eeq
where $\sigma(N_a)\equiv\sigma(i_1,i_2,\cdots,i_{N_a})$ counts 
all permutations of the $N_a$ leg labels. 
Thus 
\beqa
\Gamma^{(0)}_N &=&
\int dy_1 dy_2 \,\prod_{a=1}^3 \int dT_a
\int_{\scriptstyle x_a(T_a)=y_2 
        \atop\scriptstyle x_a(0)=y_1}[{\cal D}x_a]_{T_a}
\sum_{\sigma(N_a)}
\int_0^{T_a} d\tau^{(a)}_{i_{N_a}}
\int_0^{\tau^{(a)}_{i_{N_a}}} d\tau^{(a)}_{i_{N_a-1}}  \cdots 
\int_0^{\tau^{(a)}_{i_2}}  d\tau^{(a)}_{i_1}  \nn\\
&&\times 
\,\Bigl(\, {\tilde V}_{i_1}(\tau^{(a)}_{i_1})\cdots 
{\tilde V}_{i_{N_a}}(\tau^{(a)}_{i_{N_a}})
\Bigr)^{j_al_a}_{\mu_a\nu_a}\,\lambda(a)\ .
\eeqa

The remaining tasks to obtain the full gluon amplitudes are rather 
straightforward: insert the operators defined by \eq{vi}, which 
are at most products of two world-line fields as explained in 
Section~\ref{sec5}, and then compute correlations among these 
world-line fields by using the world-line correlator discussed 
in Appendix C (setting the extra constant background fields 
to be zero). 

\section*{Appendix C. Derivation of the background Green 
function}\label{ap3}
\setcounter{section}{3}
\setcounter{equation}{0}
\indent
In this appendix, we derive the two-loop world-line Green function 
in a constant background with demonstrating how to perform the path 
integrals in Eq.\eq{GbEH}. Let us introduce an external ($\tau$ 
dependent) source term as usual in field theory:
\beq{IJ}
z[J]\equiv\int d^Dy_1 d^Dy_2 \Bigl(\prod_{a=1}^3
\int_{\scriptstyle x_a(T_a)=y_2 
        \atop\scriptstyle x_a(0)=y_1}{\cal D}x_a\Bigr)
\exp\Bigl[S^{(b)}+
\sum_a\int_0^{T_a} J_a^\mu(\tau)x_a^\mu(\tau) d\tau \,\Bigr]\ .
\eeq
To perform the path integrals, we decompose $x_a$ into classical 
and `quantum' fields: 
\beq{*}
x_a(\tau_a) = x^c_a(\tau_a) + {\tilde x}_a(\tau_a)\ ,
\eeq
with the boundary condition ${\tilde x}_a(0)={\tilde x}_a(T_a)=0$, 
and the classical field is given by~\cite{RSS} 
\beq{class}
{x^c_a}_\mu(\tau_a)=\Bigl({y_1+y_2\over2}\Bigr)_\mu 
+R^a_{\mu\nu}(y_2-y_1)_\nu\ ,
\eeq
where 
\beq{*}
R^a_{\mu\nu}=\Bigl({e^{2i{\cal F}\tau_a}-1\over
e^{2i{\cal F}T_a}-1}-{1\over2}\Bigr)_{\mu\nu}\ .
\eeq
Then Eq.\eq{IJ} is rewritten in the form
\beqa
z[J]&=&\int d^Dy_1 d^Dy_2\prod_{a=1}^3{}^{'}
\int_{\scriptstyle {\tilde x}_a(T_a)=0 
      \atop\scriptstyle {\tilde x}_a(0)=0}{\cal D}{\tilde x}_a
\exp[\,\int_0^{T_a}\{ {1\over2}{\tilde x}_a({1\over2}\der_\tau^2 
-i{\cal F}\der_\tau){\tilde x}_a + 
J_a\cdot{\tilde x}_a\}d\tau\,]\nn\\
&&\times\exp\Bigl[\,-{1\over4}(y_2-y_1)^\mu 
A^a_{\mu\nu}(y_2-y_1)_\nu 
+(y_2-y_1)^\nu\int_0^{T_a}J_a^\mu R^a_{\mu\nu}d\tau\,\Bigr]\nn\\
&&\times\exp\Bigl[\,({y_1+y_2\over2})_\mu
\int_0^{T_a}J_a(\tau)d\tau\,\Bigr]\ ,
\eeqa
where
\beq{*}
A^a_{\mu\nu}=\Bigl({\cal F}\cot({\cal F}T_a)\Bigr)_{\mu\nu}\ .
\eeq

The path integrals (of 'quantum' fields) yield $\exp[-{1\over2}
J({1\over2}\der^2-i{\cal F}\der)^{-1}J]$, where the inverse 
operator is given~\cite{RSS} by Eq.\eq{green}, and the combination 
\eq{Gcombi} should be taken for the boundary condition 
${\tilde x}_a(T_a)={\tilde x}_a(0)=0$ as well. 
The (classical part) integrals provide the zero mode divergence 
(corresponding to the momentum conservation factor in the sense of 
Minkowski formulation) 
\beq{*}
\int d({y_1+y_2\over2})\prod_{a=1}^3{}^{'}
e^{{y_1+y_2\over2}\int J_a} = i\delta^D\Bigl(
\sum_{a=1}^3\int_0^{T_a} J_a^\mu(\tau)d\tau\Bigr)\ ,
\eeq
and determine the path integral normalization factor
\beq{*}
{\cal N}_b=(4\pi)^{D\over2}\Bigl(\prod_{a=1}^3(4\pi T_a)^{-D/2}
\Bigr)\mbox{Det}_L^{-{1\over2}}(\sum_{a=1}^3{}^{'}A^a)
\prod_a{}^{'}\mbox{Det}_L^{-{1\over2}}
[\,{\sin({\cal F}T_a)\over{\cal F}T_a}\,]\ .
\eeq

The final expression of $z[J]$ is therefore 
\beqa
z[J] &=& i\delta^D(\sum_{a=1}^3\int_0^{T_a} J_a^\mu(\tau)d\tau )
(4\pi)^{D\over2}(\prod_{a=1}^3 (4\pi T_a)^{-D\over2})
\mbox{Det}_L^{-{1\over2}}(\sum_a{}^{'}A^a) 
\prod_a{}^{'}\mbox{Det}_L^{-{1\over2}}
[\,{\sin({\cal F}T_a)\over{\cal F}T_a}\,] \nn\\ 
&&\times \exp\Bigl[\,-{1\over2}\sum_a{}^{'}\int_0^{T_a}\int_0^{T_a}
J^a_\mu(\tau_a) {\tilde G}_{\mu\nu}(\tau_a,\tau'_a) J^a_\nu(\tau'_a) 
d\tau_a d\tau'_a\,\Bigr] \nn\\
&&\times \exp\Bigl[\, (\sum_a{}^{'}A^a)^{-1}_{\rho\sigma}\,
(\,\sum_a{}^{'}\int_0^{T_a} R^a_{\rho\mu} J^a_\mu(\tau) d\tau \,)
(\,\sum_c{}^{'}\int_0^{T_c} R^c_{\sigma\nu} J^c_\nu(\tau) d\tau\,)
\,\Bigr] \ ,
\eeqa
and the two-point correlator \eq{Green} is derived as 
\beq{*}
{\cal G}_{\mu\nu}(\tau_a,\tau'_c)=
{\fder\over\fder J^\mu_a(\tau_a)}
{\fder\over\fder J^\nu_c(\tau'_c)}
\ln z[J]\Bigr|_{J=0}\ .
\eeq
%

\end{document}